\documentclass[11pt]{article}
\usepackage{amsfonts,amssymb, graphicx,a4,bm}
\newcommand{\ux}{\mathbf{x}}
\newcommand{\uy}{\mathbf{y}}
\newcommand{\uz}{\mathbf{z}}
\newcommand{\uw}{\mathbf{w}}
\newcommand{\uu}{\mathbf{u}}
\newcommand{\zero}{\mathbf{0}}
\newcommand{\Z}{\mathbb{Z}}
\newcommand{\R}{\mathbb{R}}
\newcommand{\F}{\mathfrak{F}}
\newcommand{\T}{\mathfrak{T}}
\newcommand{\cI}{\mathcal{I}}
\newcommand{\gp}{\mathfrak{p}}
\let\L=\Lambda \let\l=\lambda
\let\U=\Upsilon \let\de=\delta
\def\<{\langle} \def\>{\rangle}
\def\bbone{{\mathchoice {\rm 1\mskip-4mu l} {\rm 1\mskip-4mu l}
{\rm 1\mskip-4.5mu l} {\rm 1\mskip-5mu l}}}
\newcommand{\qed}{\hfill\rule{3mm}{3mm}}
\newtheorem{Lemma}{Lemma}
\newtheorem{Remark}{Remark}
\newtheorem{Theorem}{Theorem}
\newtheorem{Proposition}{Proposition}
\newtheorem{Conjecture}{Conjecture}

\begin{document}
\title{Clustering bounds on $n$-point correlations for unbounded spin systems}
\author{A. Abdesselam${ }^1$, A. Procacci${ }^2$,
B. Scoppola ${ }^3$}
\maketitle

\begin{center}
{\it ${ }^1$ Department of Mathematics,
P. O. Box 400137,
University of Virginia,
Charlottesville, VA 22904-4137, USA }\\
email: \texttt{malek@virginia.edu}
\end{center}

\begin{center}
{\it ${ }^2$ Departamento de Matem\'atica, Instituto de Ci\^{e}ncias Exatas,
Universidade Federal de Minas Gerais,
Av. Ant\^{o}nio Carlos, 6627 - Caixa Postal 702
30161-970 - Belo Horizonte - MG, BRASIL}\\
email: \texttt{aldo@mat.ufmg.br}
\end{center}

\begin{center}
{\it ${ }^3$ Dipartimento di Matematica, Universit\`a di Roma ``Tor Vergata'',
Via della Ricerca Scientifica - 00133 Roma, ITALY}\\
email: \texttt{scoppola@mat.uniroma2.it}
\end{center}

\bigskip

\parbox{11.8cm}{ \small
{\bf Abstract.}
We prove clustering estimates for the
truncated correlations, i.e., cumulants
of an unbounded spin system on the lattice.
We provide a unified treatment, based on cluster expansion techniques,
of four different regimes:
large mass, small interaction between sites, large self-interaction,
as well as the more delicate small self-interaction or `low temperature'
regime.
A clustering estimate in the latter regime
is needed for the Bosonic case of the
recent result obtained by Lukkarinen and Spohn
on the rigorous control on kinetic scales of quantum fluids.
}

\bigskip

\parbox{12cm}{\small
}

\tableofcontents

\section{Introduction}

In this paper we consider the following example of unbounded lattice spin
system.
Let $\L\subset {\mathbb{Z}}^d$ be a finite set of lattice sites
which contains the origin $\zero$.
For each site $\ux\in\L$ we associate a complex-valued random variable
$\psi(\ux)$ called a spin or a field. Note that we will use ``$*$''
to denote complex conjugation instead of a bar.
The collection
$\psi_\L=(\psi(\ux))_{\ux\in\L}\in\mathbb{C}^{\L}$
of these variables is sampled according to the
finite volume Gibbs measure
\begin{equation}
\<~\cdot~\>_{_\L}={1\over Z_\L}
\int_{\mathbb{C}^\L}
D\psi^* D\psi
\ e^{-H_\L(\psi_{\L})}
(\cdot)
\label{modelmeasure}
\end{equation}
where
\begin{equation}
Z_\L=
\int_{\mathbb{C}^\L}
D\psi^* D\psi
\ e^{-H_\L(\psi_{\L} )}>0
\label{Zdef}
\end{equation}
and
\[
D\psi^* D\psi= \prod_{\ux\in\L}
\frac{d \Re\psi(\ux) d\Im\psi(\ux)}{\pi}
\]
is proportional to the Lebesgue measure in $\mathbb{C}^\L$,
and the Hamiltonian with free boundary conditions is given by
\begin{equation}
H_\L(\psi_{\L})=\sum_{(\ux,\uy)\in \L^2}J(\ux-\uy)\psi^*(\ux)
\psi(\uy)+ \frac{\l}{4}\sum_{\ux\in\L}|\psi(\ux)|^4\ .
\label{Hamiltonian}
\end{equation}
The assumptions on the parameters appearing in the Hamiltonian
are the following.

\begin{itemize}
\item  $\l>0$.

\item The pair potential $J$ is a function
$\Z^d\rightarrow \mathbb{C}$ of compact support
such that
\[
J(\ux)^*=J(-\ux)\ .
\]

\item  $J_{\neq}=\sum\limits_{\ux\neq\mathbf{0}}|J(\ux)|$
is finite and moreover satisfies
$0<J_{\neq}<J(\zero)$.
\end{itemize}

Let $\psi^\sharp$ stand for either $\psi$ or $\psi^*$.
The main goal of this article is to study the so called
{\it truncated correlations functions}, defined by

\[
\<\psi^\sharp(\ux_1),\ldots,\psi^\sharp(\ux_n)\>_{\L}^{\rm T}=
\]
\[
=\left.
\frac{\partial}{\partial\alpha_1}\ldots\frac{\partial}{\partial\alpha_n}
{\rm Log}\<(1+\alpha_1\psi^\sharp(\ux_1))\ldots(1+\alpha_n\psi^\sharp(\ux_n))\>_{\L}
\right|_{\alpha_1,...,\alpha_n=0}
\]
where ${\rm Log}$ denotes the principal logarithm of a complex number, i.e.
${\rm Log}(re^{i\theta})=\ln(r)+i\theta$,
with $r>0$ and $\theta\in(-\pi,\pi]$.
It is a standard task to show that the series in $\alpha_1,...,\alpha_n$
is analytic in a small polydisc $D_\L$ around $\alpha_1,...,\alpha_n=0$.
These truncated correlations are also known in the literature as cumulants,
semi-invariants or connected correlation functions.

We will find uniformly in $\L$ and
the assignments for the $\sharp$ symbols, explicit bounds of the form
\begin{equation}
\sum_{\ux_2,\ldots,\ux_n\in\L}
|\<\psi^{\sharp_1}(\zero),\psi^{\sharp_2}(\ux_2),\ldots,\psi^{\sharp_n}(\ux_n)
\>_\L^{\rm T}|\le c_n(J,\l)
\label{clusteringsum}
\end{equation}
also called $l^1$-clustering estimates.
We will derive such bounds in four different regimes:
\begin{itemize}
\item $J(\zero)$ big, i.e., the large-mass regime;

\item $J_{\neq}$ small, i.e., the regime of small interaction between lattice
sites;

\item $\l$ big, or the `high temperature' regime;

\item $\l$ small, i.e., the near-Gaussian regime,
or the `low temperature' regime.
\end{itemize}

\begin{Remark}
This way of introducing the distinction
between high and low temperature is not what is usually done in textbook
treatments of the Ginzburg-Landau theory of phase transitions
(see e.g.~\cite[Ch. 2]{LeBellac}) which addresses models similar to the one
above. Instead, we introduce the temperature dependence by
replacing the Hamiltonian $H_\L(\psi_\L)$ by $\beta H_\L(\psi_\L)$ where
$\beta$ as usual denotes the inverse of the temperature times Boltzmann's
constant. However, one can do the change of variable
$\psi=\beta^{-\frac{1}{2}} \psi'$
and absorb $\beta$ into a modification of the coupling
$\l\rightarrow \beta^{-1}\l$. In this setting, $\l$ big coincides with
high temperature, and $\l$ small with low temperature.
In the present situation of a massive lattice model,
one does not expect a phase transition at `low temperature'.
\end{Remark}

In the statement of the theorems below
we will denote by the same symbol $O(1)$ the various
constants which appear. These are absolute constants such as $\sqrt{\pi}$
or quantities which only depend on the dimension $d$ which is fixed throughout.
In the following sections we will express such constants in details,
to prove that they are effectively computable. However we are not interested
in this
paper in optimal bounds so the values that we give may be improved
with more refined estimates.

Our results for the various regimes above consist of the following theorems.

\begin{Theorem}\label{largemassthm}
In the event that $J_{\neq}>0$ and $\l>0$ are fixed, there exists a
$K(J_{\neq},\l)>J_{\neq}>0$
which is independent of the volume $\L$ such that
the $l^1$-clustering estimate
\[
\sum_{\ux_2,\ldots,\ux_n\in\L}
|\<\psi^{\sharp_1}(\zero),\psi^{\sharp_2}(\ux_2),\ldots,\psi^{\sharp_n}(\ux_n)
\>_\L^{\rm T}|\le
n!\times\left[
\frac{O(1) }{\sqrt{J(\zero)-J_{\neq}}}
\right]^{n}
\]
holds as soon as $J(\zero)\ge K(J_{\neq},\l)$.
\end{Theorem}

\begin{Theorem}\label{smalljdiffthm}
In the event that $J(\zero)>0$ and $\l>0$ are fixed, there exists an
$\epsilon(J(\zero),\l)\in (0,J(\zero))$
which is independent of the volume $\L$ such that
the $l^1$-clustering estimate
\[
\sum_{\ux_2,\ldots,\ux_n\in\L}
|\<\psi^{\sharp_1}(\zero),\psi^{\sharp_2}(\ux_2),\ldots,\psi^{\sharp_n}(\ux_n)
\>_\L^{\rm T}|\le
n!\times\left[
\frac{O(1)
\left(1+\frac{1}{J(\zero)}\sqrt{\frac{\l}{2}}\right)
}{\sqrt{J(\zero)-J_{\neq}}}
\right]^{n}
\]
holds as soon as $0<J_{\neq}\le\epsilon(J(\zero),\l)$.
\end{Theorem}

\begin{Theorem}\label{largelambdathm}
In the event that $J_{\neq}>0$ and $J(\zero)>J_{\neq}$ are fixed,
there exists a
$K(J(\zero),J_{\neq})>0$
which is independent of the volume $\L$ such that
the $l^1$-clustering estimate
\[
\sum_{\ux_2,\ldots,\ux_n\in\L}
|\<\psi^{\sharp_1}(\zero),\psi^{\sharp_2}(\ux_2),\ldots,\psi^{\sharp_n}(\ux_n)
\>_\L^{\rm T}|\le  n!\times\left[
O(1)\times \l^{-\frac{1}{4}}
\right]^{n}
\]
holds as soon as $\l\ge K(J(\zero),J_{\neq})$.
\end{Theorem}

\begin{Theorem}\label{smalllambdathm}
Let $N\ge 1$ be an integer and
suppose that $J$ is fixed.
Then there exist quantities $\epsilon(J)>0$, $c_1(N,J)>0$
and $c_2(J)>0$ such that,
for any $\l$ with $0<\l\le \epsilon(J)$,
for any nonempty finite volume $\L\subset\Z^d$,
and any even integer $n$, $n\ge 2(N+1)$,
we have the $l^1$-clustering bound
\[
\sum_{\ux_2,\ldots,\ux_n\in\L}
|\<\psi^{\sharp_1}(\zero),\psi^{\sharp_2}(\ux_2),\ldots,\psi^{\sharp_n}(\ux_n)
\>_\L^{\rm T}|\le
c_1(N,J)\times c_2(J)^n\times \l^N\times n!\ .
\]
\end{Theorem}

\begin{Theorem}\label{twoptfunction}
There exists a quantity $c_3(J)>0$ such that
for the same $\epsilon(J)$ as in the previous theorem, for any $\l$
with $0<\l\le \epsilon(J)$,
and for any nonempty finite volume $\L\subset\Z^d$, one has
\[
\sum_{\ux_2\in\L}
|\<\psi^{\sharp_1}(\zero),\psi^{\sharp_2}(\ux_2)
\>_{\L,\l}^{\rm T}
-\<\psi^{\sharp_1}(\zero),\psi^{\sharp_2}(\ux_2)
\>_{\L,0}^{\rm T}
|\le c_3(J)\l
\]
where we have restored the $\l$-dependence in the notation for truncated
correlation functions.
\end{Theorem}

Such $l^1$-clustering estimates in statistical mechanics
have a long history. See~\cite{RuelleRMP,Ruelle}
for the case of classical gases at low activity.
The importance of such estimates in relation to analyticity of thermodynamic
functions was stressed in~\cite{DIS}.
However, to the best of our knowledge, clustering estimates which include
the results of the previous theorems have not previously appeared
in the literature.
The cluster expansion methods we used to derive these results
also provide the exponential tree decay of the truncated
correlations as a function of the locations $\ux_1,\ldots,\ux_n$
of the sources. The reader can easily extract such decay estimates
from the proofs provided here.
However, in this article we focus instead on $l^1$
estimates where the $\ux_i$ are summed over the lattice.
Indeed, our primary motivation comes from the recent work
of Lukkarinen and Spohn~\cite{LukkarinenS,LukkarinenS2} where they
obtained
the rigorous control of the kinetic regime
for the time-evolution of a class of interacting quantum fluids.
Their result is conditional on establishing $l^1$-clustering
estimates such as the ones provided in Theorem \ref{smalllambdathm}
and Theorem \ref{twoptfunction}.
Similar estimates may also be obtained in the Fermionic case in
an easier way (see~\cite{Salm2}).
Our methods are robust enough to handle much more general
undounded spin systems, but for better readability we refrained from stating
our results
with maximal generality, and restricted our attention to the
complex Bosonic model presented above which is the one
needed for~\cite{LukkarinenS,LukkarinenS2}.

Another motivation for the present work comes from the recent
interest in the decay of correlations for unbounded spin systems,
especially in relation to Log-Sobolev inequalities
(see e.g.~\cite{Zegarlinski,Yoshida,Aneetal,ProcacciS,Helffer,GuionnetZ}
and references therein).
For the equivalence between the exponential decay of the truncated
2-point function, the Log-Sobolev inequality and the spectral gap
property, in the case of
unbounded spins, see~\cite{Yoshida2}.
The study of the decay properties for the 2-point function
for this class of models has a long
history~\cite{Kunz,Gross,IsraelN,Cammarota,Sokal}.
In the low temperature case, i.e., in a regime similar to
the setting of Theorems \ref{smalllambdathm} and
\ref{twoptfunction},
recent proofs of exponential decay for the truncated 2-point function
were given in~\cite{Sjostrand1,BachJS,Sjostrand2,BachM1,Matte}.
These works use the methods introduced
by Helffer and Sj\"ostrand in~\cite{HelfferS}.
However, to the best of our knowledge, one has not been able to treat
higher truncated $n$-point functions with Helffer-Sj\"{o}strand and Witten
laplacian techniques.
The only such results~\cite{HelfferS,Lo}
for higher correlations that we are aware of
concern centered moments, i.e., expectations
of the form $\<(X_1-\<X_1\>)\ldots(X_n-\<X_n\>)\>$
which are different from fully truncated correlations
$\<X_1,\ldots,X_n\>^{\rm T}$.

It is quite well known that the first three regimes mentioned above
are amenable to cluster expansion techniques of the kind that is
standard in the statistical mechanics literature
(see e.g.~\cite{Ruelle,MalyshevM,Simon}).
Much less known, in the mathematical analysis and probability theory
communities, is that the low temperature regime
can also be treated using a special kind of cluster expansion technique.
We refer to the latter as the field theoretic cluster expansion.
It also has a long history, and originates in the work of Glimm, Jaffe and
Spencer in constructive quantum field theory~\cite{GJS1,GJS2}.
It has then been simplified and improved by many authors
(see in particular~\cite{Brydges,MackP,Pordt,GJ,Rivasseaubook}
and references therein).
Our approach for the proof of Theorems \ref{smalllambdathm}
and \ref{twoptfunction}
owes much to the pioneering work~\cite{EMS} in the context
of $P(\phi)\sb{2}$ quantum field theories.
This was adapted to the lattice setting in~\cite{Constantinescu,Wagner}.
These use the original Glimm-Jaffe-Spencer cluster expansion, which involves
a decoupling procedure for the Gaussian measure. A simpler way to do this
was introduced by Brydges, Battle and Federbush~\cite{BF,BattleF}
and it is based on a combinatorial tree expansion identity.
The third generation of such decoupling procedures, which is the one used
in this article, is based on the Brydges-Kennedy-Abdesselam-Rivasseau
(BKAR) forest formula~\cite{BK,AR1}.
We give an account of this basic tool and of the
general results of cluster expansion in \S\ref{bkarsection} and
\S\ref{clusterexp}.
We will use it many times, in the proof of the estimates,
first for the `high-temperature' regimes and then for the
`low temperature'
regime.
One of the difficulties which we have to address here and which does not
seem to have received a great deal of attention in
the previous literature, is the aim for
bounds which are uniform in $n$, growing as $n!$, and where one
{\it simultaneously} extracts as many powers of $\l$ as possible.
Such an extraction of perturbation theory in
the context of constructive field theory can be done using additional
Taylor expansions (see e.g.~\cite[\S5.14]{BIJ}).
However, we have been unable to extract the optimal bound
$C^n \l^{\frac{n}{2}-1}n!$
predicted by tree level perturbation theory, in a way which is
uniform in $n$.
The precise statement of this problem is given as
Conjecture \ref{conj} from \S\ref{treeanalysis}.
Of course, one can ask the same question for the real-valued
scalar field model with $\varphi^4$ interaction, on the lattice.
Note that a related bound with a factor $\l^{\frac{n}{4}}$ was obtained
by Brydges, Dimock and Hurd~\cite[Theorem 9]{BDH} in the context
of the UV limit
of the $\phi_3^4$ model.

Let us conclude this introduction by indicating some of the notation
used throughout this article.
We use $|\cdot|$ for the cardinality of finite sets.
If $n$ is a nonnegative integer, the set $\{1,2,\ldots,n\}$
is denoted by $[n]$. We will denote by $\bbone\{\cdots\}$
the characteristic function of the condition between braces.

\section{Preliminaries}

\subsection{Basic properties of the model}\label{basics}

By hypothesis on the function $J$, the matrix
$\tilde{J}=(J(\ux-\uy))_{\ux,\uy\in\L}$ is Hermitian positive definite.
This can be proved as follows. Let us denote the inner product on $l^2(\L)$ by
$<\psi_1,\psi_2>=\sum_{\ux\in\L} \psi_1^*(\ux)\psi_2(\ux)$, and the norm
of a vector by $||\cdot||$.
The hypothesis $J(\ux)^*=J(-\ux)$ trivially implies that the matrix
$\tilde{J}=(J(\ux-\uy))_{\ux,\uy\in\L}$ is Hermitian $\tilde{J}^\dagger=\tilde{J}$.
Besides for any field $\psi$ on $l^2(\L)$, on has
\begin{eqnarray*}
<\psi,\tilde{J}\psi> & = & \sum_{\ux,\uy\in\L}\psi^*(\ux)J(\ux-\uy)\psi(\uy)\\
 & = & J(\zero) ||\psi||^2+\sum_{{\ux,\uy\in\L}\atop{\ux\neq\uy}}
\psi^*(\ux)J(\ux-\uy)\psi(\uy)
\end{eqnarray*}
and
\[
\left|\sum_{{\ux,\uy\in\L}\atop{\ux\neq\uy}}
\psi^*(\ux)J(\ux-\uy)\psi(\uy)
\right|  \le
\sum_{{\ux,\uy\in\L}\atop{\ux\neq\uy}}
\left(|\psi(\ux)|\sqrt{|J(\ux-\uy)|}\right)
\left(|\psi(\uy)|\sqrt{|J(\ux-\uy)|}\right)
\]
\[
\le  \frac{1}{2} \sum_{{\ux,\uy\in\L}\atop{\ux\neq\uy}}
|\psi(\ux)|^2|J(\ux-\uy)|
+ \frac{1}{2} \sum_{{\ux,\uy\in\L}\atop{\ux\neq\uy}}
|\psi(\uy)|^2|J(\ux-\uy)|
\]
where we used the inequality $ab\le \frac{1}{2}(a^2+b^2)$. Therefore
the last expression is bounded by $J_{\neq}||\psi||^2$ and
\[
<\psi,\tilde{J}\psi>\ \ge (J(\zero)-J_{\neq})||\psi||^2
\]
so that $\tilde{J}$ is positive definite by the hypothesis $J(\zero)>J_{\neq}$.
The covariance matrix $C=\tilde{J}^{-1}=(C(\ux,\uy))_{\ux,\uy\in\L}$ is
well defined, and
there exists a unique mean zero
normalized Gaussian probability
measure denoted by $d\mu_C(\psi^*,\psi)$ on $\mathbb{C}^\L$ with covariance
matrix $C$, i.e., such that
\begin{eqnarray*}
\int_{\mathbb{C}^\L}\ d\mu_C(\psi^*,\psi)\
\psi(\ux) \psi^*(\uy) & = & C(\ux,\uy)\ ,\\
\int_{\mathbb{C}^\L}\ d\mu_C(\psi^*,\psi)\
\psi^*(\ux) \psi^*(\uy) & = & 0\ ,\\
\int_{\mathbb{C}^\L}\ d\mu_C(\psi^*,\psi)\
\psi(\ux) \psi(\uy) & = & 0\ .
\end{eqnarray*}
The measure can be written
\[
d\mu_C(\psi^*,\psi)=(\det\tilde{J})
e^{-\sum_{(\ux,\uy)\in \L^2}J(\ux-\uy)\psi^*(\ux)\psi(\uy)} D\psi^* D\psi\ .
\]
The moments of this measure can be expressed via the
Isserlis-Wick Theorem~\cite{Isserlis,Wick}
as follows:
\begin{itemize}
\item
If $p\neq q$ then
\[
\int_{\mathbb{C}^\L}\ d\mu_{C}(\psi^*,\psi)\
\psi(\ux_1)\ldots\psi(\ux_p) \psi^*(\uy_1)\ldots\psi^*(\uy_q)=0\ .
\]
\item
In the $p=q$ case, one has
\[
\int_{\mathbb{C}^\L}\ d\mu_C(\psi^*,\psi)\
\psi(\ux_1)\ldots\psi(\ux_p) \psi^*(\uy_1)\ldots\psi^*(\uy_p)
\]
\begin{equation}
=
\sum_{\gamma\in\mathfrak{S}_p}C(\ux_{\gamma(1)},\uy_{1})\ldots
C(\ux_{\gamma(p)},\uy_{p})\ ,
\label{IsserlisWick}
\end{equation}
i.e., one sums over all possible pairwise Wick contractions
of the $\psi$'s with the $\psi^*$'s, as indicated by the permutation
$\gamma$.
\end{itemize}

\subsection{Feynman diagrams and tree level analysis}
\label{treeanalysis}
The relation (\ref{IsserlisWick}) may be expressed in terms of oriented graphs (Feynman
diagrams) in which for each  $\psi(\ux)$ we draw  an ingoing
half edge from the
vertex $\ux$, for each  $\psi^*(\uy)$ we draw  an outgoing
half edge from the
vertex $\uy$,
and the expectation is the sum over all the possible oriented graphs obtained
contracting only outgoing half edges with ingoing half edges and
associating a free propagator $C(\ux,\uy)$ to each edge.

In this section we will focus on the near-Gaussian regime, i.e., $\l$ small,
and we will analyze the truncated $n$-point function
$\<\psi^{\sharp_1}(\ux_1),\ldots,\psi^{\sharp_n}(\ux_n)\>_{\L,\l}^{\rm T}$.
A trivial change of variable $\psi\rightarrow e^{i\theta}\psi$ shows that
truncated correlation functions
vanish unless the number of arguments $n$ is even, and there are an equal
numbers
of $\psi$'s and $\psi^*$'s involved. We will therefore always assume
this to be the case.

An easy but crucial lemma we will later need is the following.
\begin{Lemma}\label{Benedettohmwk}
For $n$ even, $n\ge 4$, and for all $k<\frac{n}{2}-1$
we have
\[
\left(\frac{{\rm d}}{{d\rm }\l}
\right)^k
\left.
\<\psi^{\sharp_1}(\ux_1),\ldots,\psi^{\sharp_n}(\ux_n)\>_{\L,\l}^{\rm T}
\right|_{\l=0}
=0\ .
\]
\end{Lemma}

We assume the reader is familiar with the rigorous formalism of Feynman
diagrams used to express formal perturbation theory. A precise mathematical
treatment can be found in~\cite{AbdesselamSLC,Salm}. A well-known fact from this
formalism is that the function
$\<\psi^\sharp(\ux_1),\ldots,\psi^\sharp(\ux_n)\>_{\L,\l}^{\rm T}$
is $C^\infty$ in $\l$ in the interval $[0,+\infty)$, and that its
Taylor series at the origin, seen as a formal power series in $\l$, is the sum
of the contributions of all connected Feynman diagrams with external legs
$\ux_1,\ldots,\ux_n$. It is a trivial exercise in organic chemistry to see
that the minimal number $N$ of internal
$4$-valent vertices needed to build such a connected graph is $\frac{n}{2}-1$.
This corresponds to tree graphs. The lemma is an immediate consequence of this
fact.

The remainder of this section will be devoted to
some heuristic considerations which we hope
will shed some light on the near-Gaussian regime.
It is part of constructive field theory folklore
that one should expect the correct bound in the $l^1$-clustering estimate
to be dictated by the contribution of tree graphs.
For a given tree graph, the $l^1$ sum over the sites $\ux_2,\ldots,\ux_n$
in $\Z^d$ can be easily bounded, using a standard pin and sum argument,
by ${\rm Cst}^n$.
This is because the hypotheses on $J$ imply the exponential decay of the free
propagator $C(\ux,\uy)$, as is recalled in \S\ref{freedecay}.
The issue is the number of trees.
Because of the constraints due to the orientations of the edges, the counting
is not an immediate consequence of Cayley's formula. One can nevertheless
obtain an exact formula.

For $n\ge 2$ and even,
let $\kappa_{\frac{n}{2}}$ denote the number of Wick
contractions $\gamma$, with $n$ external legs and
$N=\frac{n}{2}-1$ internal vertices, which produce connecting trees.
One can easily check by inspection that $\kappa_1=1$, $\kappa_2=4$,
$\kappa_3=288$,
and $\kappa_4=82944$. In general one has the following result.

\begin{Lemma}\label{exactcount}
For any $k\ge 1$ one has
\[
\kappa_k=\frac{2^{k-1}k!(k-1)!(3k-3)!}{(2k-1)!}\ .
\]
\end{Lemma}

\noindent{\bf Proof:}
We find a quadratic induction formula for the $\kappa_k$,
namely, for any $k\ge 3$
one has
\[
(k-2)\kappa_k=\sum\limits_{i=1}^{k-2}
\left(
\begin{array}{c}
k-1\\
i
\end{array}
\right)
\left(
\begin{array}{c}
k\\
i+1
\end{array}
\right)
\left(
\begin{array}{c}
k\\
k-i
\end{array}
\right)
\kappa_{i+1} \kappa_{k-i}\ .
\]
Indeed, the left-hand side counts Wick contraction schemes together with
the choice of an internal edge of the corresponding trees.
If one cuts that distinguished edge, the tree falls apart into two trees $T_1$,
$T_2$. The numbering is unambiguous, if one decides that the cut edge
goes from $T_2$ to $T_1$.
Let $i$, $1\le i\le k-2$, be the number of internal vertices
of $T_1$. Choosing them accounts for the first binomial coefficient.
The second binomial coefficient is the number of ways one can choose the
$i+1$ external $\psi$ vertices of $T_1$ among the inital $k=\frac{n}{2}$.
The third coefficient is for the choices of the $k-i$ external $\psi^*$
vertices in $T_2$. Note that the cut edge introduces an extra distinguished
$\psi^*$ leaf for $T_1$ and an extra $\psi$ leaf for $T_2$.
Now the given formula for $\kappa_k$ satisfies the quadratic induction
because of the identity
\begin{equation}
\sum\limits_{j=0}^{k-1}
\frac{(3j)!(3k-3j-3)!}{j!(2j+1)!(2k-2j-1)!(k-j-1)!}
=\frac{(3k-2)!}{k!(2k-1)!}
\label{combident}
\end{equation}
which holds for any $k\ge 1$, and which is an easy consequence
of~\cite[Eq. 5.62]{Knuth}.
\qed

\begin{Remark}
One can also prove (\ref{combident}) using Clausen's ${}_4 F_3$
hypergeometric summation formula.
The $\kappa_k$'s are related to the Fuss-Catalan numbers of order 3
(see~\cite[p. 347]{Knuth}). 
\end{Remark}

Now the rough estimate for the $l^1$-clustering bound which comes
from the aforementioned analysis of the tree level
contribution, for even $n\ge 2$,
is
\[
c_n(\l)\sim
{\rm Cst}^n \l^{\frac{n}{2}-1}\frac{\kappa_{\frac{n}{2}}}{N!}\sim
{\rm Cst}^n \l^{\frac{n}{2}-1}n!
\]
by Stirling's formula and Lemma \ref{exactcount}.

It is therefore natural to make the following conjecture.

\begin{Conjecture}\label{conj}
The exists a constant $c(J)>0$,
such that for $\l>0$ small enough,
for any even integer $n\ge 2$, for any finite volume $\L\in\Z^d$,
and for any assignment of the $\sharp$ symbols,
one has
\[
\sum_{\ux_2,\ldots,\ux_n\in\L}
|\<\psi^{\sharp_1}(\zero),\psi^{\sharp_2}(\ux_2),\ldots,\psi^{\sharp_n}(\ux_n)
\>_{\L,\l}^{\rm T}|\le c(J)^n \l^{\frac{n}{2}-1}n!\ .
\]
\end{Conjecture}

Although we do not see a fundamental reason against this conjecture,
due to technical difficulties inherent to the cluster expansion
method we follow in this paper, we have been unable to prove so much.
Related bounds of the $n!$ type have been obtained in the
literature~\cite{EMS,Pordt,BDH},
but without extracting the optimal power of $\l$.
Using the method of this article, it is possible to extract this power,
but at the cost of a higher power of $n!$.
The difficulty is in obtaining optimal bounds which are {\em uniform} in $n$.
Theorem \ref{smalllambdathm} is a weakening of this conjecture.

\subsection{Free propagator decay}\label{freedecay}

Picking up the thread from \S\ref{basics}, we
let
$\tilde{J}_{\neq}$ denote the off-diagonal part of $\tilde{J}$, so that
$\tilde{J}=J(\zero) {\rm I}+\tilde{J}_{\neq}$. The matrix $\tilde{J}_{\neq}$
is also Hermitian and, because of the previous inequalities, has its operator
norm bounded by $||\tilde{J}_{\neq}||\le J_{\neq}<J(\zero)$.
Thus the Neumann series
\[
C=\frac{1}{J(\zero)}\sum_{p\ge 0}
\left(\frac{-1}{J(\zero)}\tilde{J}_{\neq}\right)^p
\]
converges, and provides the following random path representation for the free
propagator
\[
C(\ux,\uy) =  \frac{1}{J(\zero)}\delta_{\ux,\uy}
-\frac{1}{J(\zero)^2}\bbone_{\{\ux\neq\uy\}}
J(\ux-\uy)
+\sum\limits_{p\ge 2}\frac{(-1)^p}{J(\zero)^{p+1}}
\]
\begin{equation}
\times
\sum_{\uz_1,\ldots,\uz_{p-1}} J(\ux-\uz_1)J(\uz_1-\uz_2)\cdots J(\uz_{p-2}-\uz_{p-1})
J(\uz_{p-1}-\uy)
\label{randompath}
\end{equation}
where the last sum is over sequences of sites in $\L$ such that, $\uz_1\neq\ux$,
$\uz_{p-1}\neq\uy$, and $\uz_{i}\neq\uz_{i-1}$ for $2\le i\le p-1$.
We assumed the function $J:\Z^d\rightarrow\mathbb{C}$ to be compactly supported.
Let therefore $r_0>0$ be such that $J(\ux)=0$ for all $\ux\in\Z^d$
satisfying $|\ux|\ge r_0$.
We now have the following elementary lemma.

\begin{Lemma}\label{decaylemma}
For any $\L\subset\Z^d$, and for any $\ux,\uy\in\L$
one has the uniform exponential decay bound
\[
|C(\ux,\uy)|\le K_0 e^{-\mu_0|\ux-\uy|}
\]
where
\[
K_0=\frac{J(\zero)}{J_{\neq}(J(\zero)-J_{\neq})}>0
\]
and
\[
\mu_0=\frac{1}{r_0} \log\left(
\frac{J(\zero)}{J_{\neq}}
\right)\ .
\]
\end{Lemma}

\noindent{\bf Proof:}
Given a nonempty $\L$ in $\Z^d$, and $\ux,\uy$ in $\L$, let
$p_0=\lfloor \frac{|\ux-\uy|}{r_0}\rfloor$.
First suppose that $p_0\ge 1$, so that $|\ux-\uy|\ge r_0$.
Then obviously the first two terms on the right hand side of
(\ref{randompath}) vanish. Besides, if a term in the last sum over $p\ge 2$ is
nonzero, then there exists a sequence of sites $\uz_1,\ldots,\uz_{p-1}$
such that $|\ux-\uz_1|<r_0$, $|\uz_{p-1}-\uy|<r_0$, and $|\uz_i-\uz_{i-1}|<r_0$ for
$2\le i\le p-1$. Thus $|\ux-\uy|<pr_0$ and $p_0<p$, i.e., $p\ge p_0+1$.
Therefore at the level of matrix elements on has
\[
C(\ux,\uy)=\left[
\frac{1}{J(\zero)}\sum_{p\ge p_0+1}
\left(\frac{-1}{J(\zero)}\tilde{J}_{\neq}\right)^p
\right]_{\ux,\uy}
\]
and bounding matrix elements by operator norms, one has
\[
|C(\ux,\uy)|\le \frac{1}{J(\zero)}
\left(
\frac{J_{\neq}}{J(\zero)}
\right)^{p_0+1}\times
\frac{1}{1-\frac{J_{\neq}}{J(\zero)}}
\]
from which the estimate easily follows, since $p_0+1>\frac{|\ux-\uy|}{r_0}$.
Now if $p_0=0$, one has $|\ux-\uy|<r_0$ and therefore using the
full Neumann series and the coarse bound by the operator norm
\[
|C(\ux,\uy)|\le ||C||\le \frac{1}{J(\zero)-J_{\neq}}
=  \frac{1}{J(\zero)-J_{\neq}} e^{- \mu_0|\ux-\uy|}\times e^{\mu_0|\ux-\uy|}
\]
and the last factor is obviously bounded by
$e^{\mu_0 r_0}=\frac{J(\zero)}{J_{\neq}}$, so the lemma
follows.
\qed

This exponential decay trivially implies the $l^1$ property
\begin{equation}
\sum_{\uz\in\Z^d} e^{-\mu|\uz|}\le K_1(d,\mu)
\label{expdecayeq}
\end{equation}
for any $\mu>0$ and dimension $d\ge 1$, where $K_1(d,\mu)$
is some constant.

\subsection{The Brydges-Kennedy-Abdesselam-Rivasseau formula}\label{bkarsection}
We recall in this  section and the next one a
list of well known combinatorial identities and inequalities
and some basic results of cluster expansion that
we will use to establish our bounds.

The BKAR formula~\cite{BK,AR1} is a simpler
and more symmetric version of the earlier Brydges-Battle-Federbush
tree formula~\cite{BF,BattleF,Brydges} in constructive field theory.
This earlier formula itself is an improvement on the pioneering approach
of Glimm-Jaffe-Spencer~\cite{GJS1,GJS2}.

Let us consider a finite set $E\neq\emptyset$, and let us denote by $E^{(2)}$
the set unordered pairs $\{a,b\}$, where $a$ and $b$ are any
distinct
elements
in $E$. Of course $|E^{(2)}|=
\left(
\begin{array}{c}
|E|\\
2
\end{array}
\right)$.
We will consider the space $\R^{E^{(2)}}$ of multiplets
$s=(s_l)_{l\in E^{(2)}}$
indexed by pairs $l\in E^{(2)}$, and functions defined on a particular
compact convex set $\mathcal{K}_E$ in this space.
Let $\Pi_E$ denote the set of partitions of $E$. For any
partition $\pi=\{X_1,\ldots
X_q\}$ in $\Pi_E$ we associate a vector $v_\pi=(v_{\pi,l})_{l\in E^{(2)}}$
defined as
\[
v_{\pi,l}=\bbone_{\{\exists i, 1\le i\le q, l\subset X_i\}}\ .
\]
Now $\mathcal{K}_E$ is by definition the convex hull of the vectors
$v_\pi$, for $\pi\in\Pi_E$.
It is easy to see that $\mathcal{K}_E$ affinely generates
$\R^{E^{(2)}}$. Indeed, let $\hat{0}$ be the partition entierly made
of singletons, and for any pair $l\in E^{(2)}$ let $\hat{l}$
denote the partition made of the two element set $l$ and the
singletons
$\{a\}$, for $a\in E\backslash l$.
Then, the vectors $v_{\hat{l}}-v_{\hat{0}}$, for $l\in E^{(2)}$
form a basis of the vector space $\R^{E^{(2)}}$.
As a result, the open domain $\Omega_E=\mathring{\mathcal{K}}_E$
is nonempty, and $\mathcal{K}_E$ is equal to the closure $\bar{\Omega}_E$.
Let $C^{k}(\bar{\Omega}_E)$ denote the usual space of functions of class
$C^k$ on the domain $\Omega_E$ which, together with their derivatives
up to order $k$, admit uniformly continuous extentions to the closure
$\mathcal{K}_E=\bar{\Omega}_E$ (see, e.g.,~\cite{AdamsF}).

Now a simple graph with vertex set $E$ can be thought of as a
subset of the complete graph $E^{(2)}$.
A forest $\F$ is a graph with no circuits, and it is made of a
vertex-disjoint collection of trees.
Let $\F$ be a forest, and let $\vec{h}=(h_l)_{l\in\F}$
be a vector of real parameters indexed by the edges $l$ in the forest $\F$.
To such data we canonically associate a multiplet
$s(\F,\vec{h})=(s(\F,\vec{h})_l)_{l\in E^{(2)}}$ in $\R^{E^{(2)}}$
as follows.
Let $a$ and $b$ be two distinct elements in $E$.
If $a$ and $b$ belong to two distinct connected components
of the forest $\F$, then $s(\F,\vec{h})_{\{a,b\}}=0$.
Otherwise let, by definition, $s(\F,\vec{h})_{\{a,b\}}=\min\limits_l h_l$
where $l$ belongs to the unique path in the forest $\F$
joining $a$ to $b$.
We are now ready to state the BKAR formula.

\begin{Theorem}~\cite{BK,AR1}
\label{bkarthm}
Let $f\in C^{|E|-1}(\bar{\Omega}_E)$, and let $1\in \R^{E^{(2)}}$
denote the multiplet with all entries equal to one.
This is also the same as $v_{\hat{1}}$ where
$\hat{1}$ is the single block partition $\{E\}$.
We then have
\[
f(1)=
\sum_{\F\ {\rm forest}}\int_{[0,1]^\F}\ d\vec{h}\
\frac{\partial^{|\F|}f}{\prod_{l\in\F}\partial s_l}\left(
s(\F,\vec{h})
\right)
\]
where the sum is over all forests $\F$ with vertex set $E$,
the notation $d\vec{h}$ is for the Lebesgue measure on the set
of parameters $[0,1]^\F$,
the partial derivatives of $f$ are with respect to the entries
indexed by the pairs belonging to $\F$, and the evaluation of these
derivatives is at the $\vec{h}$ dependent point $s(\F,\vec{h})$.
Such points belong to $\mathcal{K}_E$.
\end{Theorem}

Note that the empty forest always occurs and its contribution
is $f(0)=f(v_{\hat{0}})$.
There are several proofs of this identity~\cite{BK,AR1,BM},
but we believe the most natural
and most easily generalizable is the one given in~\cite[\S2]{AR2}.
This proof also most clearly shows the $s(\F,\vec{h})$ belong to
$\mathcal{K}_E$. This point is important for the positivity of the interpolated
covariance matrices in \S\ref{GJSsection},
and also when proving Lemma \ref{treeinequality} via
Lemma \ref{exptreeformula}.

We will now recall a lemma which,
via the uniqueness of the M\"{o}bius inverse in the partition lattice,
is a corollary of the BKAR forest formula.

\begin{Lemma}\label{exptreeformula}
Again let us consider a finite set $E$ and let us denote by $E^{(2)}$
the set of unordered pairs $l=\{a,b\}$ in $E$. Let $V_{\{a,b\}}$ be a collection
of complex numbers indexed by $E^{(2)}$.
Then
\begin{equation}
\sum_{\mathsf{g}\leadsto E} \prod_{l\in\mathsf{g}}
\left(
e^{-V_l}-1
\right)
=
\sum_{{\mathfrak{T}\leadsto E}\atop{\mathfrak{T}\ {\rm tree}}}
\int_{[0,1]^{\mathfrak{T}}}\ d\vec{h}\ \
\left\{\prod_{l\in\mathfrak{T}} (-V_l)\right\}
\ e^{-\sum_{l\in E^{(2)}}  s(\T,\vec{h})_l V_l}\ .
\label{graphvstree}
\end{equation}
Here $\mathsf{g}$ is summed over all simple graphs (i.e. subsets of $E^{(2)}$)
which connect $E$. We abreviate this property by
the notation $\mathsf{g}\leadsto E$.
On the right-hand side the sum is on spanning trees
$\mathfrak{T}$ which connect $E$. The notation $s(\mathfrak{T},\vec{h})$
is as in Theorem \ref{bkarthm}.
\end{Lemma}

The following tree graph inequality, initially
due to Brydges, Battle and Federbush (see ~\cite{Brydges,BattleF,PdLS})
is the basic tool we will need for the estimates in the first three
regimes related to the `high temperature' scenario.
It is an easy consequence of Lemma \ref{exptreeformula}.

\begin{Lemma}\label{treeinequality}
Under the same hypotheses as in Lemma \ref{exptreeformula}, let us assume
that the numbers $V_l$ satisfy, in addition, the following stability hypothesis:
there are nonnegative numbers $U_a$, for $a\in E$, such that for any subset
$S\subset E$ one has
\[
\left|\sum_{l\in S^{(2)}}V_l\right|\le\sum_{a\in S}U_a\ .
\]
Then the following inequality holds
\[
\left|
\sum_{\mathsf{g}\leadsto E} \prod_{l\in\mathsf{g}}
\left(
e^{-V_l}-1
\right)
\right|\le
e^{\sum_{a\in E}U_a}
\sum_{{\mathfrak{T}\leadsto E}\atop{\mathfrak{T}\ {\rm tree}}}
\prod_{l\in\mathfrak{T}} |V_l|\ .
\]
\end{Lemma}

\subsection{The cluster expansion for the polymer gas}\label{clusterexp}
We now recall the basics of polymer gas cluster expansions.
Any nonempty finite subset $R\subset\L$ is called a polymer. We denote by
$\mathbf{P}(\L)$ the set of all such polymers. We associate to each
$R\in \mathbf{P}(\L)$ a variable $\rho(R)\in \mathbb{C}$ called
the activity of the the polymer $R$.
They can be collected in a vector
$\rho=(\rho(R))_{R\in\mathbf{P}(\L)}\in\mathbb{C}^{\mathbf{P}(\L)}$.

On the complex space $\mathbb{C}^{\mathbf{P}(\L)}$
we consider the polynomial function $\mathcal{Z}$
defined by
\[
\mathcal{Z}(\rho)=
\sum_{p\ge 0}\frac{1}{p!}
\sum_{R_1,\ldots,R_p\in \mathbf{P}(\L)}
\bbone\left\{
\begin{array}{c}
{\rm the}\ R_i\ {\rm are}\\
{\rm disjoint}
\end{array}
\right\}
\rho(R_1)\ldots\rho(R_p)
\]
for any $\rho\in\mathbb{C}^{\mathbf{P}(\L)}$.
This function is usually called the {\it grand canonical
partition function
of the polymer gas} at finite volume $\L$.
It is well known (see e.g. \cite{Brydges, Cammarota}) that the
logarithm of $\mathcal{Z}_\L$ can be written in terms
of the following series
\[
\log \mathcal{Z}(\rho)=\sum_{p\ge 1}\frac{1}{p!}
\sum_{R_1,\ldots,R_p\in \mathbf{P}(\L)}
\phi^{\rm T}(R_1,\ldots,R_p)
\rho(R_1)\ldots\rho(R_p)
\]
with
\[
\phi^{\rm T}(R_1,\ldots,R_p)=\sum_{{H\leadsto[p]}\atop{H\subset G}}
(-1)^{|H|}
\]
where $G$ is the graph with vertex set $[p]=\{1,\ldots,p\}$
and edges corresponding to the pairs $\{i,j\}$, $i\neq j$, such that
$R_i\cap R_j\neq \emptyset$.
The sum is over all spanning connecting subgraphs $H$
which are identified with their edge sets.

Note that by the so-called Whitney-Tutte-Fortuin-Kasteleyn
representation (see, e.g.,~\cite{Sokal2}), one can write the
chromatic polynomial of $G$ as
\begin{equation}
P(G,x)=\sum_{H\subset G}(-1)^{|H|}\ x^{c(H)}
\label{WTFKrep}
\end{equation}
where $c(H)$ is the number of connected components of $H$.
For nonnegative integer values of $x$, the quantity $P(G,x)$ is by definition
the number of proper vertex colorings of $G$ with $x$ colors.
A good way to see the Ursell function $\phi^{\rm T}(R_1,\ldots,R_p)$
is as the coefficient of $x$ in the chromatic polynomial $P(G,x)$.

The condition for the convergence of the series above is a well
studied subject. It  can be expressed in terms of the following norm, depending on a parameter $a>0$,
defined
on the space $\mathbb{C}^{\mathbf{P}(\L)}$ of polymer activities
\[
||\rho||_a=\sup_{\ux\in\L}\sum_{R\in\mathbf{P}(\L)}
\bbone\{\ux\in R\}\ |\rho(R)|\ e^{a|R|}\ .
\]

The best result on this subject, essentially proven almost four
decades ago by Gruber and
Kunz \cite{GruberK}  but largely forgotten, and then rediscovered very
recently by Fern\'andez
and Procacci  \cite{FernandezP} with a new proof, is the following
theorem.
\begin{Theorem}\label{abstractlog}
Let  $a>0$
and let $\rho$ denote an element of $\mathbb{C}^{\mathbf{P}(\L)}$.
Then the series
\[
f(\rho)=\sum_{p\ge 1}\frac{1}{p!}
\sum_{R_1,\ldots,R_p\in \mathbf{P}(\L)}
\phi^{\rm T}(R_1,\ldots,R_p)
\rho(R_1)\ldots\rho(R_p)
\]
is absolutely convergent in the closed ball $||\rho||_a\le e^a-1$.
The function $f$ is analytic on the open ball $||\rho||_a<e^a-1$
and satisfies
\[
\exp f(\rho)=\mathcal{Z}
\]
for any $\rho$ with $||\rho||_a\le e^a-1$.
\end{Theorem}

In fact, one can extract a more precise result
(see~\cite[p. 132]{FernandezP}), when $\rho\ge 0$, i.e., when the
polymer activities $\rho(R)$ are real and nonnegative.

\begin{Theorem}
\label{Pibound}
If $\rho\ge 0$ and $||\rho||_a\le e^a-1$, then for any $R_0\in \mathbf{P}(\L)$,
we have the estimate
\[
\sum_{p\ge 1}\frac{1}{p!}
\sum_{R_1,\ldots,R_p\in \mathbf{P}(\L)}
|\phi^{\rm T}(R_0,R_1,\ldots,R_p)|.
\rho(R_1)\ldots\rho(R_p)
\le e^{a|R_0|}-1\ .
\]
\end{Theorem}

We will use this theorem when $R_0$ is a singleton, namely, when
$R_0=\{\uz\}$ for some $\uz\in\L$.
In this case one has the identity
\begin{equation}
\phi^{\rm T}(R_0,R_1,\ldots,R_p)=(-r).\phi^{\rm T}(R_1,\ldots,R_p)
\label{ursellreduct}
\end{equation}
where $r$ is the number of indices $q$, $1\le q\le p$,
such that $\uz\in R_q$.
Indeed, if $G$ is the intersection graph on $\{0,1,\ldots,p\}$
defined by the collection of polymers $R_0,R_1,\ldots,R_p$,
then the restriction of $G$ to the set formed by the vertex $0$
and its neighbors is the complete graph on $r+1$ elements.
This property and the representation (\ref{WTFKrep}) gives a very easy proof
of the reduction formula (\ref{ursellreduct}).
A trivial consequence of (\ref{ursellreduct}) and Theorem~\ref{Pibound}
is the following lemma which will be used repeatedly in the sequel.

\begin{Lemma}
\label{pinsumlemma}
For nonnegative polymer activities $\rho(R)$, $R\in\mathbf{P}(\L)$,
such that $||\rho||_a\le e^a-1$ we have the bound
\begin{eqnarray}
\lefteqn{
\sup\limits_{\uz\in\L}
\sum_{p\ge 1}\frac{1}{p!}
\sum_{R_1,\ldots,R_p\in \mathbf{P}(\L)}
\bbone\{\uz\in\cup_{q=1}^p R_q\}
} & & \nonumber\\
 & & \times |\phi^{\rm T}(R_1,\ldots,R_p)|.
\rho(R_1)\ldots\rho(R_p)
\le e^a-1\ .
\label{totalpinsum}
\end{eqnarray}
\end{Lemma}

\begin{Remark}
The important point to note here is that the bound is uniform in $\L\in\Z^d$.
\end{Remark}

Since we are not interested in optimal bounds we will
choose hereafter $a=\log 2$ and we will denote the norm
$||\rho||_{\log 2}$ simply by $||\rho||$.
So in what follows we will use the norm
\begin{equation}\label{norm}
||\rho||=\sup_{\ux\in\L}\sum_{R\in\mathbf{P}(\L)}
\bbone\{\ux\in R\}\ |\rho(R)|\ 2^{|R|}\
\end{equation}
with the condition ensuring absolute convergence of $f(\rho)$ being
\begin{equation}\label{conditconv}
||\rho||\le 1\ .
\end{equation}

\section{The large mass, small interaction, and large self-interaction regimes}

\subsection{The Mayer series representation for the truncated correlations}
\label{mayercorrelsec}

Given a source specification $(\ux_i,\sharp_i)_{i\in I}$,
we will consider the perturbed partition function
\[
Z_\L(\bm{\alpha})=
\int_{\mathbb{C}^\L}
D\psi^* D\psi
\ e^{-H_\L(\psi_{\L} )}
\ \prod_{i\in I} (1+\alpha_i \psi^{\sharp_i}(\ux_i))
\]
and the truncated correlation function given by
\[
\<(\psi^{\sharp_i}(\ux_i))_{i\in I}\>_{\L}^{\rm T}=
\left.\frac{\partial^{|I|}}{\prod_{i\in I} \partial\alpha_i}
\ {\rm Log}\ Z_\L(\bm{\alpha})\right|_{\bm{\alpha}=\mathbf{0}}\ .
\]
As mentioned before, ${\rm Log}\ Z_\L(\bm{\alpha})$ is analytic in a small
polydisc $D_\L$ around $\bm{\alpha}=\mathbf{0}$.

Now let us introduce the normalized single site measure
\[
d\nu(z^*,z)=\frac{1}{\mathcal{N}}
e^{-J(\zero)|z|^2-\frac{\l}{4}|z|^4}
d\Re z\ d\Im z
\]
on $\mathbb{C}$, where
\[
\mathcal{N}=\int_{\mathbb{C}}
e^{-J(\zero)|z|^2-\frac{\l}{4}|z|^4}
d\Re z\ d\Im z\ .
\]
Clearly $Z_\L(\bm{\alpha})=\mathcal{N}^{|\L|}\check{Z}_{\L}(\bm{\alpha})$
where
\[
\check{Z}_{\L}(\bm{\alpha})=
\int_{\mathbb{C}^\L}
\prod_{\ux\in\L} d\nu(\psi^*(\ux),\psi(\ux))
\ e^{-\sum_{\{\ux,\uy\}\in \L^{(2)}} I_{\{\ux,\uy\}}(\psi)}
\ \prod_{i\in I} (1+\alpha_i \psi^{\sharp_i}(\ux_i))
\]
with
\[
I_{\{\ux,\uy\}}(\psi)=J(\ux-\uy)\psi^*(\ux)\psi(\uy)+
J(\uy-\ux)\psi^*(\uy)\psi(\ux)\ .
\]
Since  $Z_\L(\bm{\alpha})=\mathcal{N}^{|\L|}\check{Z}_{\L}(\bm{\alpha})$
with $\mathcal{N}>0$,
one has the analyticity in the polydisc $D_\L$
of ${\rm Log}\ \check{Z}_\L(\bm{\alpha})=-|\L|\ln \mathcal{N}+
{\rm Log}\ Z_\L({\bm{\alpha}})$
as well as the identity
\[
\<(\psi^{\sharp_i}(\ux_i))_{i\in I}\>_{\L}^{\rm T}=
\left.\frac{\partial^{|I|}}{\prod_{i\in I} \partial\alpha_i}
\ {\rm Log}\ \check{Z}_\L(\bm{\alpha})\right|_{\bm{\alpha}=\zero}\ .
\]

On the other hand, we can also write
\begin{eqnarray*}
e^{-\sum_{\{\ux,\uy\}\in \L^{(2)}} I_{\{\ux,\uy\}}(\psi)} & = &
\prod_{\{\ux,\uy\}\in\L^{(2)}}
\left[1+(e^{-I_{\{\ux,\uy\}}(\psi)}-1)\right]\\
 & = & \sum_{\mathsf{g}\subset\L^{(2)}}
\prod_{\{\ux,\uy\}\in\mathsf{g}}
\left(e^{-I_{\{\ux,\uy\}}(\psi)}-1\right)
\end{eqnarray*}
where the sum is over all simple graphs $\mathsf{g}$ on the vertex set $\L$.
Using this equation and also expanding the product of the
$(1+\alpha_i \psi^{\sharp_i}(\ux_i))$ one easily obtains, after reorganization
according to the connected components of the graph $\mathsf{g}$,
the following polymer
representation for $\check{Z}_{\L}(\bm{\alpha})$.
Namely, one has
\begin{equation}
\check{Z}_{\L}(\bm{\alpha})=
\sum_{p\ge 0}\frac{1}{p!}
\sum_{R_1,\ldots,R_p\in\mathbf{P}(\L)}
\bbone\left\{
\begin{array}{c}
{\rm the}\ R_i\ {\rm are}\\
{\rm disjoint}
\end{array}
\right\}
\rho(R_1,\bm{\alpha})\ldots
\rho(R_p,\bm{\alpha})
\label{polyrep}
\end{equation}
where
the polymer activity of a polymer $R$ is defined as
\begin{eqnarray}
\lefteqn{
\rho(R,\bm{\alpha})=
\sum_{\mathsf{g}\leadsto R}
\sum_{J\subset I_R}
\bbone\left\{
\begin{array}{c}
|R|\ge 2\\
{\rm or}\ J\neq\emptyset
\end{array}
\right\}
} & & \nonumber\\
 & & \int_{\mathbb{C}^R}
\prod_{\ux\in R} d\nu(\psi^*(\ux),\psi(\ux))
\prod_{i\in J}
\left(\alpha_i \psi^{\sharp_i}(\ux_i)\right)
\prod_{\{\ux,\uy\}\in\mathsf{g}}
\left(e^{-I_{\{\ux,\uy\}}(\psi)}-1\right)
\label{rhoalpha}
\end{eqnarray}
with the notation $I_R=\{i\in I| \ux_i\in R\}$.
Let $\rho(\bm{\alpha})$ denote
the activity specification $(\rho(R,\bm{\alpha}))_{R\in\mathbf{P}(\L)}$.
Using the notation and definitions of \S\ref{clusterexp}
one obviously has
$\check{Z}_{\L}(\bm{\alpha})=\mathcal{Z}(\rho(\bm{\alpha}))$.

Now suppose the condition $||\rho(\mathbf{0})||<1$ holds.
Then on a small polydisc $D'_\L$ the hypothesis $||\rho(\bm{\alpha})||<1$
will also hold.
On the polydisc $D_\L\cap D'_\L$ both the functions
$f(\rho(\bm{\alpha}))$ and ${\rm Log}\ \check{Z}_{\L}(\bm{\alpha})$
are analytic and exponentiate to $\check{Z}_{\L}(\bm{\alpha})$.
By connectedness the difference
$f(\rho(\bm{\alpha}))-{\rm Log}\ \check{Z}_{\L}(\bm{\alpha})$
is a constant in $2i\pi\mathbb{Z}$.
However this difference takes a real value at $\bm{\alpha}=\mathbf{0}$,
and therefore it must vanish.
Indeed, the hypothesis on the $J(\ux)$ function implies that
$I_{\{\ux,\uy\}}(\psi)$ is real and therefore, by (\ref{rhoalpha}),
the activities $\rho(R,\mathbf{0})$ are also real.
Finally, by the definition in Theorem \ref{abstractlog}
it follows that $f(\rho(\mathbf{0}))\in\mathbb{R}$.
As a result of these considerations, on a small polydisc around
$\bm{\alpha}= \mathbf{0}$ one has the equality
\[
f(\rho(\bm{\alpha}))={\rm Log}\ \check{Z}_{\L}(\bm{\alpha})
\]
thus
\begin{eqnarray*}
\<(\psi^{\sharp_i}(\ux_i))_{i\in I}\>_{\L}^{\rm T} & = & \left.
\frac{\partial^{|I|}}{\prod_{i\in I} \partial\alpha_i}
f(\rho(\bm{\alpha}))
\right|_{\bm{\alpha}=\zero}\\
 & = & \sum_{p\ge 0}\frac{1}{p!}
\sum_{R_1,\ldots,R_p\in\mathbf{P}(\L)}
\phi^{\rm T}(R_1,\ldots,R_p)\\
 & & \times
\left.\frac{\partial^{|I|}}{\prod_{i\in I} \partial\alpha_i}
\rho(R_1,\bm{\alpha})\ldots
\rho(R_p,\bm{\alpha})
\right|_{\bm{\alpha}=\zero}\\
 & = & \sum_{p\ge 0}\frac{1}{p!}
\sum_{R_1,\ldots,R_p\in\mathbf{P}(\L)}
\phi^{\rm T}(R_1,\ldots,R_p)
\sum_{I_1,\ldots,I_p\subset I}
\bbone\left\{
\begin{array}{c}
I_q\ {\rm disjoint}\\
\cup I_q=I
\end{array}
\right\}\\
 & & \times\prod_{q=1}^p \left[
\left.\frac{\partial^{|I_q|}}{\prod_{i\in I_q} \partial\alpha_i}
\rho(R_q,\bm{\alpha})
\right|_{\bm{\alpha}=\zero}
\right]\ .
\end{eqnarray*}
Comparing with (\ref{rhoalpha}) it is apparent that the effect of the
$\alpha$ derivatives is to force the summed over $J$ to be equal to $I_q$.
Hence, as an immediate consequence of Theorem \ref{abstractlog},
we have the following statement.

\begin{Proposition}\label{correlseries}
Define for any subset $J$ of the source label set $I$,
and any polymer $R\in\mathbf{P}(\L)$,
\[
\tilde{\rho}(R,J)=
\bbone\left\{
\begin{array}{c}
\forall i\in J\\
\ux_i\in R
\end{array}
\right\}
\bbone\left\{
\begin{array}{c}
|R|\ge 2\\
{\rm or}\ J\neq\emptyset
\end{array}
\right\}
\sum_{\mathsf{g}\leadsto R}
\]
\[
\int_{\mathbb{C}^R}
\prod_{\ux\in R} d\nu(\psi^*(\ux),\psi(\ux))
\prod_{i\in J}
\psi^{\sharp_i}(\ux_i)
\prod_{\{\ux,\uy\}\in\mathsf{g}}
\left(e^{-I_{\{\ux,\uy\}}(\psi)}-1\right)\ .
\]
Provided the source-free condition
\[
\sup_{\ux\in\L}\sum_{R\in\mathbf{P}(\L)}
\bbone\{\ux\in R\}\ |\tilde{\rho}(R,\emptyset)|\ 2^{|R|}<1
\]
holds, one has the absolutely convergent series representation
for all truncated correlation functions
\begin{eqnarray*}
\lefteqn{
\<(\psi^{\sharp_i}(\ux_i))_{i\in I}\>_{\L,\l}^{\rm T}=} & & \\
 & & \sum_{p\ge 1}\frac{1}{p!}
\sum_{{R_1,\ldots,R_p\in\mathbf{P}(\L)}\atop{I_1,\ldots,I_p\subset I}}
\phi^T(R_1,\ldots,R_p)
\bbone\left\{
\begin{array}{c}
I_q\ {\rm disjoint}\\
\cup I_q=I
\end{array}
\right\}
\prod_{q=1}^p \tilde{\rho}(R_q,I_q)\ .
\end{eqnarray*}
\end{Proposition}
Note that with this new definition, polymer activities
$\tilde{\rho}(R,J)$
do not have to contain all the sources localized in sites belonging to $R$.

\subsection{Estimates for a single polymer activity}

Consider a polymer activity $\tilde{\rho}(R,J)$
as defined in Proposition \ref{correlseries}.
By moving the sum over connecting graphs inside the integral,
one has the estimate
\[
|\tilde{\rho}(R,J)|\le
\bbone\left\{
\begin{array}{c}
\forall i\in J\\
\ux_i\in R
\end{array}
\right\}
\bbone\left\{
\begin{array}{c}
|R|\ge 2\\
{\rm or}\ J\neq\emptyset
\end{array}
\right\}
\int_{\mathbb{C}^R}
\prod_{\ux\in R} d\nu(\psi^*(\ux),\psi(\ux))
\]
\[
\left(
\prod_{i\in J}|\psi^{\sharp_i}(\ux_i)|
\right)\times
\left|
\sum_{\mathsf{g}\leadsto R}
\prod_{\{\ux,\uy\}\in\mathsf{g}}
\left(e^{-I_{\{\ux,\uy\}}(\psi)}-1\right)
\right|\ .
\]
Now for any subset $S\subset R$ and for any fixed field $\psi$, by
the argument given in \S\ref{basics}, one has
\[
\left|\sum_{\{\ux,\uy\}\in S^{(2)}}I_{\{\ux,\uy\}}(\psi)\right|
\le J_{\neq}\sum_{\ux\in S} |\psi(\ux)|^2\ .
\]
As result, Lemma \ref{treeinequality} implies, for any given field
configuration $\psi$, the inequality
\begin{eqnarray*}
\lefteqn{
\left|
\sum_{\mathsf{g}\leadsto R}
\prod_{\{\ux,\uy\}\in\mathsf{g}}
\left(e^{-I_{\{\ux,\uy\}}(\psi)}-1\right)
\right|\le} & & \\
 & & e^{J_{\neq}\sum_{\ux\in R} |\psi(\ux)|^2}
\sum_{{\mathfrak{T}\leadsto R}\atop{\mathfrak{T}\ {\rm tree}}}
\prod_{\{\ux,\uy\}\in\mathfrak{T}}
\left(2|J(\ux-\uy)||\psi(\ux)||\psi(\uy)|
\right)\ .
\end{eqnarray*}
Therefore
\[
|\tilde{\rho}(R,J)|\le
\bbone\left\{
\begin{array}{c}
\forall i\in J\\
\ux_i\in R
\end{array}
\right\}
\bbone\left\{
\begin{array}{c}
|R|\ge 2\\
{\rm or}\ J\neq\emptyset
\end{array}
\right\}\ 2^{|R|-1}
\sum_{{\mathfrak{T}\leadsto R}\atop{\mathfrak{T}\ {\rm tree}}}
\prod_{\{\ux,\uy\}\in\mathfrak{T}} |J(\ux-\uy)|
\]
\begin{equation}
\times\prod_{\ux\in R}
\int_{\mathbb{C}}
d\nu(\psi^*(\ux),\psi(\ux))
e^{J_{\neq}|\psi(\ux)|^2} |\psi(\ux)|^{m(\ux)}
\label{rawmassive}
\end{equation}
with $m(\ux)=c_J(\ux)+d_{\mathfrak{T}}(\ux)$
where $c_J(\ux)$ is number of source labels $i\in J$ such that
$\ux_i=\ux$, while
$d_{\mathfrak{T}}(\ux)$ is the degree of the vertex $\ux$ in the tree
$\mathfrak{T}$.
The integrals in (\ref{rawmassive})
are now estimated thanks to the following lemma.

\begin{Lemma}
For any integer $m\ge 0$, and under the hypotheses of the Introduction, we
have
\begin{equation}
\int_{\mathbb{C}}
d\nu(z^*,z)
e^{J_{\neq}|z|^2} |z|^{m}
\le
2\left(J(\zero)+\sqrt{\frac{\l}{2}}\right)
(J(\zero)-J_{\neq})^{-\frac{m+2}{2}}
m!^{\frac{1}{2}}
\label{option1}
\end{equation}
as well as
\begin{equation}
\int_{\mathbb{C}}
d\nu(z^*,z)
e^{J_{\neq}|z|^2} |z|^{m}
\le
4\left(J(\zero)+\sqrt{\frac{\l}{2}}\right)
(\frac{\l}{4})^{-\frac{m+2}{4}}
m!^{\frac{1}{4}}\ .
\label{option2}
\end{equation}
\end{Lemma}

\noindent{\bf Proof:}
We have
\[
\int_{\mathbb{C}}
d\nu(z^*,z)
e^{J_{\neq}|z|^2} |z|^{m}=\frac{1}{\mathcal{N}}
\int_{\mathbb{C}}
e^{-(J(\zero)-J_{\neq})|z|^2-\frac{\l}{4}|z|^4}\ |z|^m
\ d\Re z\ d\Im z\ .
\]
In order to obtain a lower bound on the denominator $\mathcal{N}$, we write
\[
\mathcal{N}  =  \int_{\mathbb{C}}
e^{-J(\zero)|z|^2-\frac{\l}{4}|z|^4}
d\Re z\ d\Im z
=  2\pi \int\limits_0^\infty
e^{-J(\zero)r^2-\frac{\l}{4}r^4} r dr
=  \pi  \int\limits_0^\infty
e^{-\frac{\l}{4}t^2-J(\zero)t}dt
\]
\[
=  \pi e^{\frac{J(\zero)^2}{\l}}
\int\limits_0^\infty
e^{-\frac{1}{2}\left(\sqrt{\frac{\l}{2}}t+J(\zero)\sqrt{\frac{2}{\l}}\right)^2}dt
=  \pi\sqrt{\frac{2}{\l}}
e^{\frac{J(\zero)^2}{\l}}
\int\limits_{J(\zero)\sqrt{\frac{2}{\l}}}^\infty
e^{-\frac{q^2}{2}} dq\ .
\]

Now recall Birnbaum's inequality~\cite{Birnbaum} for Mill's ratio, i.e.,
essentially the ${\rm erfc}$ function:
\[
e^{\frac{x^2}{2}}\int\limits_x^\infty e^{-\frac{q^2}{2}} dq
\ge \frac{\sqrt{4+x^2}-x}{2}
\]
for $x\ge 0$.
One can simplify this to
\[
e^{\frac{x^2}{2}}\int\limits_x^\infty e^{-\frac{q^2}{2}} dq
\ge
\frac{2}{x+\sqrt{4+x^2}}\ge
\frac{2}{x+\sqrt{4+4x+x^2}}=\frac{1}{x+1}\ .
\]
The latter applied to $x=J(\zero)\sqrt{\frac{2}{\l}}$
readily provides the needed bound
\begin{equation}
\mathcal{N}\ge\frac{\pi}{J(\zero)+\sqrt{\frac{\l}{2}}}\ .
\end{equation}
Now for the upper bound on the numerator, we have two possibilities
corresponding to the two estimates in the lemma.

\noindent{\bf 1st option--the bound using the Gaussian part:}
We write
\[
\int_{\mathbb{C}}
e^{-(J(\zero)-J_{\neq})|z|^2-\frac{\l}{4}|z|^4}\ |z|^m
\ d\Re z\ d\Im z
\le
\int_{\mathbb{C}}
e^{-(J(\zero)-J_{\neq})|z|^2}\ |z|^m
\ d\Re z\ d\Im z
\]
\[
=2\pi\int\limits_0^\infty
e^{-(J(\zero)-J_{\neq})r^2} r^{m+2} \frac{dr}{r}
=2\pi(J(\zero)-J_{\neq})^{-\frac{m+2}{2}}\Gamma\left(\frac{m+2}{2}\right)\ .
\]
We therefore only need to show the elementary inequality
$\frac{1}{\sqrt{m!}}\Gamma\left(\frac{m+2}{2}\right)\le 1$
in order to complete the proof of (\ref{option1}).
This can be done easily by induction using the well-known properties
of Euler's gamma function (see e.g.~\cite{AndrewsAR}).
The quantity $\frac{1}{\sqrt{m!}}\Gamma\left(\frac{m+2}{2}\right)$
is equal to $1$ and $\frac{\sqrt{\pi}}{2}<1$ for
$m=0$, and $m=1$ respectively. Besides, when going from $m$ to
$m+2$, this quantity changes by a factor of
\[
\frac{1}{\sqrt{(m+1)(m+2)}}\times\frac{m+2}{2}=\frac{1}{2}
\sqrt{1+\frac{1}{m+1}}<1
\]
and the desired inequality propagates.

\noindent{\bf 2nd option--the bound using the quartic self-interaction:}
We write
\[
\int_{\mathbb{C}}
e^{-(J(\zero)-J_{\neq})|z|^2-\frac{\l}{4}|z|^4}\ |z|^m
\ d\Re z\ d\Im z
\le
\int_{\mathbb{C}}
e^{-\frac{\l}{4}|z|^4}\ |z|^m
\ d\Re z\ d\Im z
\]
\[
=2\pi\int\limits_0^\infty
e^{-\frac{\l}{4}r^4} r^{m+2} \frac{dr}{r}
=2\pi\left(\frac{\l}{4}\right)^{-\frac{m+2}{4}}\Gamma
\left(\frac{m+2}{4}\right)\ .
\]
We are left with showing $\Gamma\left(\frac{m+2}{4}\right)\le 2 m!^{\frac{1}{4}}$
for any nonnegative integer $m$.
By Dirichlet's multidimensional extension of the formula for the
beta integral~\cite[Theorem 1.8.6]{AndrewsAR}, we have
\begin{eqnarray*}
\Gamma\left(\frac{m+2}{4}\right)^4 & = &
(m+1)! \int_{t_i>0,\sum t_i=1}(t_1 t_2 t_3 t_4)^{\frac{m-2}{4}} \ dt_1 dt_2 dt_3 dt_4\\
 & \le & (m+1)! \int_{t_i>0,\sum t_i=1} \left(\frac{1}{4^4}\right)^{\frac{m-2}{4}}
\ dt_1 dt_2 dt_3 dt_4
\end{eqnarray*}
by the arithmetic vs. geometric mean inequality.
Thus
\[
\Gamma\left(\frac{m+2}{4}\right)^4  \le
(m+1)!\times\frac{1}{3!}\times 4^{-m+2}
\le  2^4 m!
\]
for $m\ge 0$. Indeed,
\[
4^m=(1+3)^m\ge 1+3m\ge \frac{(m+1)}{6}\ .
\]
\qed

From the raw estimate (\ref{rawmassive}), the previous Lemma, the inequality
\[
m(\ux)!\le 2^{c_J(\ux)+d_{\mathfrak{T}}(\ux)}
c_J(\ux)! d_{\mathfrak{T}}(\ux)!
\]
and the trivial relations
\[
\sum_{\ux\in R} c_J(\ux)=|J| \qquad , \qquad
\sum_{\ux\in R} d_{\mathfrak{T}}(\ux)= 2|R|-2
\]
one easily derives the following basic bounds on single polymer activities.

\begin{Lemma} {\bf Gaussian estimate:}\label{lemmaopt1}

One has the bound
\[
|\tilde{\rho}(R,J)|\le
\bbone\left\{
\begin{array}{c}
\forall i\in J\\
\ux_i\in R
\end{array}
\right\}
\bbone\left\{
\begin{array}{c}
|R|\ge 2\\
{\rm or}\ |J|\ge 1
\end{array}
\right\}\ 2^{3|R|+\frac{|J|}{2}-2}
\left(J(\zero)+\sqrt{\frac{\l}{2}}\right)^{|R|}
\]
\[
\times
(J(\zero)-J_{\neq})^{-2|R|-\frac{|J|}{2}+1}
\times \prod_{\ux\in R} c_J(\ux)!^{\frac{1}{2}}
\times
\sum_{{\mathfrak{T}\leadsto R}\atop{\mathfrak{T}\ {\rm tree}}}
\prod_{\ux\in R} d_{\mathfrak{T}}(\ux)!^{\frac{1}{2}}
\prod_{\{\ux,\uy\}\in\mathfrak{T}} |J(\ux-\uy)|\ .
\]
\end{Lemma}

\begin{Lemma} {\bf Quartic estimate (a.k.a. domination):}\label{lemmaopt2}

One has the bound
\[
|\tilde{\rho}(R,J)|\le
\bbone\left\{
\begin{array}{c}
\forall i\in J\\
\ux_i\in R
\end{array}
\right\}
\bbone\left\{
\begin{array}{c}
|R|\ge 2\\
{\rm or}\ |J|\ge 1
\end{array}
\right\}\ 2^{\frac{11|R|}{2}+\frac{3|J|}{4}-\frac{5}{2}}
\left(J(\zero)+\sqrt{\frac{\l}{2}}\right)^{|R|}
\]
\[
\times
\l^{-|R|-\frac{|J|}{4}+\frac{1}{2}}
\times \prod_{\ux\in R} c_J(\ux)!^{\frac{1}{4}}
\times
\sum_{{\mathfrak{T}\leadsto R}\atop{\mathfrak{T}\ {\rm tree}}}
\prod_{\ux\in R} d_{\mathfrak{T}}(\ux)!^{\frac{1}{4}}
\prod_{\{\ux,\uy\}\in\mathfrak{T}} |J(\ux-\uy)|\ .
\]
\end{Lemma}

\subsection{The large mass and small interaction regimes}
\label{largemasssection}

In this section we will use the estimate of Lemma \ref{lemmaopt1}
which allows us to treat at once both the cases when $J(\zero)$ is large
or when $J_{\neq}$ is small.
Our first task is to check the source-free (i.e. $J=\emptyset$)
condition for the applicability
of Proposition \ref{correlseries}.
Let $\uz$ be a site in $\L$.
We need to bound
\begin{eqnarray*}
A & = & \sum_{R\in\mathbf{P}(\L)}
\bbone\{\uz\in R\}\ |\tilde{\rho}(R,\emptyset)|\ 2^{|R|}\\
 & \le & \sum_{R\in\mathbf{P}(\L)}
\bbone\{\uz\in R\}\bbone\{|R|\ge 2\}
2^{4|R|-2}
\left(J(\zero)+\sqrt{\frac{\l}{2}}\right)^{|R|}\\
 & & \times
(J(\zero)-J_{\neq})^{-2|R|+1}
\times
\sum_{{\mathfrak{T}\leadsto R}\atop{\mathfrak{T}\ {\rm tree}}}
\prod_{\ux\in R} d_{\mathfrak{T}}(\ux)!^{\frac{1}{2}}
\prod_{\{\ux,\uy\}\in\mathfrak{T}} |J(\ux-\uy)|\ .
\end{eqnarray*}
We now condition this sum on the size $m=|R|\ge 2$ of the polymer,
and introduce a sum over labelings of the sites in $R$ by the fixed
set of indices $[m]$.
Namely we introduce in the sum the identity
\begin{equation}
1=\frac{1}{(m-1)!}\sum_{\ux_1,\ldots,\ux_m\in\L}
\bbone\{\ux_1=\uz\}
\bbone\{\ux_i\ {\rm distinct}\}
\bbone\{R=\{\ux_1,\ldots,\ux_m\}\}\ .
\label{relabeling}
\end{equation}
The next step is to use this artifice to transport the summation over
trees $\mathfrak{T}$
on the variable set $R$ to a sum over trees $\mathfrak{t}$ on the fixed set
$[m]$.
Then, partly releasing the second condition in (\ref{relabeling}),
and eliminating $R$ we obtain
\[
A\le
\sum_{m\ge 2}
\frac{1}{(m-1)!}
\sum_{\ux_1,\ldots,\ux_m\in\L}
\bbone\{\ux_1=\uz\}
2^{4m-2}
\left(J(\zero)+\sqrt{\frac{\l}{2}}\right)^{m}
\]
\[
\times
(J(\zero)-J_{\neq})^{-2m+1}
\times
\sum_{{\mathfrak{t}\leadsto [m]}\atop{\mathfrak{t}\ {\rm tree}}}
\prod_{i\in [m]} d_{\mathfrak{t}}(i)!^{\frac{1}{2}}
\prod_{\{i,j\}\in\mathfrak{t}}
\left(\bbone\{\ux_i\neq\ux_j\}
|J(\ux_i-\ux_j)|\right)
\]
with the obvious notation $d_{\mathfrak{t}}(i)$
for the degree of $i\in [m]$ in the tree
$\mathfrak{t}$.
Now the sum over the locations $\ux_i$, starting with the
leafs and then progressing towards the root $1\in [m]$,
is easy and gives a factor $J_{\neq}^{m-1}$.
The sum over the tree is done using the following easy lemma.

\begin{Lemma}\label{sumtreedegree}
We have the bound
\[
\sum_{{\mathfrak{t}\leadsto [m]}\atop{\mathfrak{t}\ {\rm tree}}}
\prod_{i\in [m]} d_{\mathfrak{t}}(i)!
\le 2^{3m-3} (m-2)!\ .
\]
\end{Lemma}

\noindent{\bf Proof:}
We have by Cayley's Theorem (see, e.g,~\cite[Theorem 5.3.4]{Stanley2})
which counts labelled spanning trees with fixed vertex degrees
\begin{eqnarray*}
\sum_{{\mathfrak{t}\leadsto [m]}\atop{\mathfrak{t}\ {\rm tree}}}
\prod_{i=1}^m d_{\mathfrak{t}}(i)!
 & = & \sum_{{d_1,\ldots,d_m\ge 1}\atop{\Sigma d_i=2m-2}}
\sum_{{\mathfrak{t}\leadsto [m]}\atop{\mathfrak{t}\ \rm tree}}
\bbone\left\{
\begin{array}{c}
\forall i,\ {\rm degree}\ {\rm of}\\
i\ {\rm in}\ \mathfrak{t}\ {\rm is}\ d_i
\end{array}
\right\}
\prod_{i=1}^m d_{\mathfrak{t}}(i)!\\
 & = & \sum_{{d_1,\ldots,d_m\ge 1}\atop{\Sigma d_i=2m-2}}
\frac{(m-2)!}{\prod_{i=1}^{m}(d_i-1)!}
\prod_{i=1}^m d_i!\\
 & = & (m-2)!\sum_{{d_1,\ldots,d_m\ge 1}\atop{\Sigma d_i=2m-2}} d_1\ldots d_m\ .
\end{eqnarray*}
Using the arithmetic versus geometric mean inequality
\[
d_1\ldots d_m\le \left[\frac{2m-2}{m}\right]^{m}\le 2^m
\]
as well as
\[
\sum_{{d_1,\ldots,d_m\ge 1}\atop{\Sigma d_i=2m-2}} 1=
\left(
\begin{array}{c}
2m-3\\
m-1
\end{array}
\right)\le 2^{2m-3}
\]
the lemma follows.
\qed

Thanks to the coarse bound
$d_{\mathfrak{t}}(i)!^{\frac{1}{2}}\le d_{\mathfrak{t}}(i)!$ and the last lemma,
we now have
\[
A\le
\sum_{m\ge 2}
2^{7m-5}
J_{\neq}^{m-1}
\left(J(\zero)+\sqrt{\frac{\l}{2}}\right)^{m}
(J(\zero)-J_{\neq})^{-2m+1}\ .
\]
Therefore, as soon as the condition
\begin{equation}
\frac{2^7 J_{\neq}.\left(J(\zero)+\sqrt{\frac{\l}{2}}\right)}
{(J(\zero)-J_{\neq})^2}
\le\frac{1}{2}
\label{cvcriterion1}
\end{equation}
holds, one will have
\begin{equation}
||\rho(\cdot,\emptyset)||\le A\le
\frac{2^{10} J_{\neq}.\left(J(\zero)+\sqrt{\frac{\l}{2}}\right)^2}
{(J(\zero)-J_{\neq})^3}\ .
\end{equation}
Clearly, if we take either one of the limits $J(\zero)\rightarrow \infty$
or $J_{\neq}\rightarrow 0$, the condition (\ref{cvcriterion1}) will hold and
$||\rho(\cdot,\emptyset)||$ can be made arbitrarily small.
Note that crucial to this last fact is $m\ge 2$, i.e., the absence of
single-site polymers, also called monomers.

We are now in a position to tackle the $l^1$-clustering estimate
(\ref{clusteringsum}), where the source label set is $I=[n]$.
Assuming either $J(\zero)$ is large enough or $J_{\neq}$ is small enough
to garantee conditions (\ref{cvcriterion1}) and
$||\rho(\cdot,\emptyset)||<1$ hold, we can use
Proposition \ref{correlseries} as well as Lemma \ref{lemmaopt1}
to write
\begin{eqnarray}
\lefteqn{
\sum_{\ux_2,\ldots,\ux_n\in\L}
|\<\psi^{\sharp_1}(\zero),\psi^{\sharp_2}(\ux_2),\ldots,\psi^{\sharp_n}(\ux_n)
\>_\L^{\rm T}| } & & \nonumber\\
 & & \le
\sum_{\ux_1,\ldots,\ux_n\in\L}\bbone\{\ux_1=\zero\}
\sum_{p\ge 1}\frac{1}{p!}
\sum_{{R_1,\ldots,R_p\in\mathbf{P}(\L)}\atop{I_1,\ldots,I_p\subset [n]}}\nonumber\\
 & & |\phi^T(R_1,\ldots,R_p)|
\bbone\left\{
\begin{array}{c}
I_q\ {\rm disjoint}\\
\cup I_q=[n]
\end{array}
\right\}
\prod\limits_{q=1}^p\left[
\bbone\left\{
\begin{array}{c}
\forall i\in I_q\\
\ux_i\in R_q
\end{array}
\right\}
\bbone\left\{
\begin{array}{c}
|R_q|\ge 2\\
{\rm or}\ |I_q|\ge 1
\end{array}
\right\}
\right.\nonumber\\
 & & \times
2^{3|R_q|+\frac{|I_q|}{2}-2}
\left(J(\zero)+\sqrt{\frac{\l}{2}}\right)^{|R_q|}
(J(\zero)-J_{\neq})^{-2|R_q|-\frac{|I_q|}{2}+1}\nonumber\\
 & & \left. \times
\prod_{\ux\in R_q} c_{I_q}(\ux)!^{\frac{1}{2}}
\times
\sum_{{\mathfrak{T}_q\leadsto R_q}\atop{\mathfrak{T}_q\ {\rm tree}}}
\prod_{\ux\in R_q} d_{\mathfrak{T}_q}(\ux)!^{\frac{1}{2}}
\prod_{\{\ux,\uy\}\in\mathfrak{T}_q} |J(\ux-\uy)|
\right]\ .
\label{monster1}
\end{eqnarray}
The first step is to push $\bbone\{\ux_1=\zero\}$ through the sums over
$p$, the $R_q$'s and the $I_q$'s. We then bound it by the coarser condition
$\bbone\{\zero\in\cup_{q=1}^p R_q\}$.
The second step is to push the sums over the $\ux_i$'s
inside the appropriate (i.e., as dictated by the choice of the $I_q$'s)
bracket factor.
The sums over the source localizations $\ux_i$ are therefore bounded
with the knowledge of which polymer they belong to.
This rests on the following lemma.

\begin{Lemma}\label{suminpoly}
Using the notation $c_{\uy}(\ux)=|\{j,1\le j\le k| \uy_j=\ux\}|$,
for
any polymer $R$, and for any power $\beta$,
we have
\[
\sum_{\uy_1,\ldots,\uy_k\in R}\
\prod_{\ux\in R} c_{\uy}(\ux)!^\beta\le
2^{|R|+k-1} \times k!^{\max(\beta,1)}\ .
\]
\end{Lemma}

\noindent{\bf Proof:}
Summing first over multiindices $c=c(\ux)_{\ux\in R}$ and then
over the sequences $(\uy_1,\ldots,\uy_k)$ for which $c_{\uy}=c$,
we
have
\[
\sum_{\uy_1,\ldots,\uy_k\in R}
\ \prod_{\ux\in R} c_{\uy}(\ux)!^\beta
=k!\sum_{c,|c|=k}
\ \prod_{\ux\in R} c(\ux)!^{\beta-1}
\]
where we denoted by $|c|$ the length $\sum_{\ux\in R}c(\ux)$
of the multiindex $c$.
We then bound the product by $k!^{\beta-1}$ if $\beta\ge 1$ and by $1$
otherwise. We also use the trivial bound
\[
\sum_{c,|c|=k} 1=\left(
\begin{array}{c}
|R|-1+k\\
|R|-1
\end{array}
\right)\le 2^{|R|+k-1}
\]
and the result follows.
\qed

As a result of this lemma we have for every $q$, $1\le q\le p$,
\begin{equation}
\sum_{(\ux_i)_{i\in I_q}\in \L^{I_q}}
\bbone\left\{
\begin{array}{c}
\forall i\in I_q\\
\ux_i\in R_q
\end{array}
\right\}
\times \prod_{\ux\in R_q} c_{I_q}(\ux)!^{\frac{1}{2}}
\le
2^{|R_q|+|I_q|-1}\times |I_q|!\ .
\label{localsumhalf}
\end{equation}
Calling $\mathfrak{W}$ the sum to be estimated on the left-hand side of
(\ref{monster1}), and using the relation $\sum_{q=1}^p |I_q|=n$ in order
to pull out some factors from the sums,
we now have
\begin{eqnarray*}
\mathfrak{W} & \le & 2^{\frac{3n}{2}}. (J(\zero)-J_{\neq})^{-\frac{n}{2}}
\sum_{p\ge 1}\frac{1}{p!}
\sum_{{R_1,\ldots,R_p\in\mathbf{P}(\L)}\atop{I_1,\ldots,I_p\subset [n]}}
\bbone\{\zero\in\cup_{q=1}^p R_q\}\\
 & & \times
|\phi^T(R_1,\ldots,R_p)|
\bbone\left\{
\begin{array}{c}
I_q\ {\rm disjoint}\\
\cup I_q=[n]
\end{array}
\right\}\times |I_1|!\times\cdots\times|I_p|!\\
 & & \times\prod\limits_{q=1}^p\left[
\bbone\left\{
\begin{array}{c}
|R_q|\ge 2\\
{\rm or}\ |I_q|\ge 1
\end{array}
\right\}
2^{4|R_q|-3} \left(J(\zero)+\sqrt{\frac{\l}{2}}\right)^{|R_q|}\right.\\
 & & \left. \times
(J(\zero)-J_{\neq})^{-2|R_q|+1}
\times
\sum_{{\mathfrak{T}_q\leadsto R_q}\atop{\mathfrak{T}_q\ {\rm tree}}}
\prod_{\ux\in R_q} d_{\mathfrak{T}_q}(\ux)!^{\frac{1}{2}}
\prod_{\{\ux,\uy\}\in\mathfrak{T}_q} |J(\ux-\uy)|
\right]\ .
\end{eqnarray*}

We now insert the decomposition
\begin{equation}
\bbone\left\{
\begin{array}{c}
|R_q|\ge 2\\
{\rm or}\ |I_q|\ge 1
\end{array}
\right\}
=
\bbone\{|R_q|\ge 2\}+
\bbone\left\{
\begin{array}{c}
|R_q|=1 \\
{\rm and}\ |I_q|\ge 1
\end{array}
\right\}
\label{dichotomy}
\end{equation}
so each bracket factor takes the form $A_q+B_q$ where the nonnegative
numbers $A_q$ and $B_q$ correspond to the first and second conditions of
(\ref{dichotomy}) respectively.
Note that since the $I_q$'s form a disjoint decomposition with possibly
empty subsets of the set $[n]$,
the number of $q$'s for which $B_q\neq 0$ is bounded by $n$.
Hence, for any number $\gamma$ such that $0<\gamma\le 1$, we can write
\begin{eqnarray*}
\prod_{q=1}^p (A_q+B_q) & = & \prod_{{q=1}\atop{B_q\neq 0}}^p (A_q+B_q)
\times \prod_{{q=1}\atop{B_q=0}}^p A_q\\
 & \le & \prod_{{q=1}\atop{B_q\neq 0}}^p (\gamma^{-1} A_q+ B_q)
\times \prod_{{q=1}\atop{B_q=0}}^p A_q\\
 & \le & \gamma^{-|\{q,B_q\neq 0\}|}
\times \prod_{{q=1}\atop{B_q\neq 0}}^p (A_q+\gamma B_q)
\times \prod_{{q=1}\atop{B_q=0}}^p A_q\\
 & \le & \gamma^{-n}\times \prod_{q=1}^p (A_q+\gamma B_q)\\
 & \le & \gamma^{-n}\times \prod_{q=1}^p (A_q+\gamma \tilde{B}_q)
\end{eqnarray*}
where $\tilde{B}_q$ means we now forget about the condition $|I_q|\ge 1$.
Leaving the appropriate choice of $\gamma$ for later,
we now have
\begin{eqnarray*}
\mathfrak{W} & \le & \gamma^{-n}. 2^{\frac{3n}{2}}. (J(\zero)-J_{\neq})^{-\frac{n}{2}}
\sum_{p\ge 1}\frac{1}{p!}
\sum_{{R_1,\ldots,R_p\in\mathbf{P}(\L)}\atop{I_1,\ldots,I_p\subset [n]}}
\bbone\{\zero\in\cup_{q=1}^p R_q\}\\
 & & \times
|\phi^T(R_1,\ldots,R_p)|
\bbone\left\{
\begin{array}{c}
I_q\ {\rm disjoint}\\
\cup I_q=[n]
\end{array}
\right\}\times |I_1|!\times\cdots\times|I_p|!\\
 & & \times\prod\limits_{q=1}^p\left[
\bbone\left\{
|R_q|\ge 2
\right\}
2^{4|R_q|-3} \left(J(\zero)+\sqrt{\frac{\l}{2}}\right)^{|R_q|}\right.\\
 & & \times
(J(\zero)-J_{\neq})^{-2|R_q|+1}
\times
\sum_{{\mathfrak{T}_q\leadsto R_q}\atop{\mathfrak{T}_q\ {\rm tree}}}
\prod_{\ux\in R_q} d_{\mathfrak{T}_q}(\ux)!^{\frac{1}{2}}
\prod_{\{\ux,\uy\}\in\mathfrak{T}_q} |J(\ux-\uy)|\\
 & & \left. \ \ + \gamma \bbone\{|R_q|=1\}\times 2
 \left(J(\zero)+\sqrt{\frac{\l}{2}}\right)\times (J(\zero)-J_{\neq})^{-1}
\right]\ .
\end{eqnarray*}

We now perform the sum over the $I_q$'s
\begin{eqnarray}
\lefteqn{
\sum_{I_1,\ldots,I_p\subset [n]}
\bbone\left\{
\begin{array}{c}
I_q\ {\rm disjoint}\\
\cup I_q=[n]
\end{array}
\right\}\times |I_1|!\times\cdots\times|I_p|! } & & \nonumber\\
 &  & = \sum_{{k_1,\ldots,k_p\ge 0}\atop{k_1+\cdots+k_p=n}}
\left(
\begin{array}{c}
n \\
k_1\ \cdots\ k_p
\end{array}
\right)
\times k_1!\ldots k_p!\nonumber\\
 &  & = n! \left(
\begin{array}{c}
n+p-1\\
p-1
\end{array}
\right)\nonumber\\
 &  & \le n! 2^{n+p-1}\ .
\label{sumis}
\end{eqnarray}

Thus
\begin{eqnarray}
\mathfrak{W} & \le & n!\ .\gamma^{-n}. 2^{\frac{5n}{2}-1}.
(J(\zero)-J_{\neq})^{-\frac{n}{2}}
\times\sum_{p\ge 1}\frac{1}{p!}
\sum_{R_1,\ldots,R_p\in \mathbf{P}(\L)}\nonumber\\
 & &
\bbone\{\zero\in\cup_{q=1}^p R_q\}
|\phi^{\rm T}(R_1,\ldots,R_p)|.
\varrho(R_1)\ldots\varrho(R_p)
\label{Wready}
\end{eqnarray}
where the nonnegative polymer activities $\varrho(\cdot)$ are defined by
\begin{eqnarray*}
\varrho(R) & = & 4\gamma.\bbone\{|R|=1\}\times
\frac{J(\zero)+\sqrt{\frac{\l}{2}}}{J(\zero)-J_{\neq}}\\
 & & +\bbone\left\{|R|\ge 2\right\}
2^{4|R|-2} \left(J(\zero)+\sqrt{\frac{\l}{2}}\right)^{|R|}
(J(\zero)-J_{\neq})^{-2|R|+1}\\
& & \ \times
\sum_{{\mathfrak{T}\leadsto R}\atop{\mathfrak{T}\ {\rm tree}}}
\prod_{\ux\in R} d_{\mathfrak{T}}(\ux)!^{\frac{1}{2}}
\prod_{\{\ux,\uy\}\in\mathfrak{T}} |J(\ux-\uy)|\ .
\end{eqnarray*}

The norm $||\varrho||$ is estimated in the same manner as we did for
$||\rho(\cdot,\emptyset)||$ at the beginning of this section.
Namely, we find
\begin{eqnarray*}
||\varrho|| & \le & 8\gamma\times
\frac{J(\zero)+\sqrt{\frac{\l}{2}}}{J(\zero)-J_{\neq}}\\
 & & +\sum_{m\ge 2}
2^{8m-5}
J_{\neq}^{m-1}
\left(J(\zero)+\sqrt{\frac{\l}{2}}\right)^{m}
(J(\zero)-J_{\neq})^{-2m+1}\ .
\end{eqnarray*}
By taking $J(\zero)$ large or $J_{\neq}$ small we will ensure that
\[
\frac{2^8 J_{\neq}.\left(J(\zero)+\sqrt{\frac{\l}{2}}\right)}
{(J(\zero)-J_{\neq})^2}
\le\frac{1}{2}\ .
\]
and also, summing the geometric series, that
\begin{equation}
||\varrho||  \le  8\gamma\times
\frac{J(\zero)+\sqrt{\frac{\l}{2}}}{J(\zero)-J_{\neq}}
+\frac{2^{12} J_{\neq}.\left(J(\zero)+\sqrt{\frac{\l}{2}}\right)^2}
{(J(\zero)-J_{\neq})^3}
\label{finetuning}
\end{equation}
holds.

\noindent{\bf 1st case:} When $J(\zero)$ becomes large, the right-hand side of
(\ref{finetuning})
approaches $8\gamma$. So we simply choose $\gamma=\frac{1}{16}<1$
and Lemma \ref{pinsumlemma} together with the previous estimate
(\ref{Wready}) complete the proof of Theorem \ref{largemassthm}.

\noindent{\bf 2nd case:} When $J_{\neq}$ becomes small,
the right-hand side of
(\ref{finetuning})
approaches
$8\gamma \left(1+\frac{1}{J(\zero)}\sqrt{\frac{\l}{2}}\right)$.
Therefore, we choose
\[
\gamma =
\frac{1}{16 \left(1+\frac{1}{J(\zero)}\sqrt{\frac{\l}{2}}\right)}
<1
\]
and now Theorem \ref{smalljdiffthm} follows.

\subsection{The large self-interaction regime}

This essentially is an encore presentation of Section \ref{largemasssection}.
The difference is that we now use the estimate of Lemma \ref{lemmaopt2}.
Following the same line of argument as in \S\ref{largemasssection},
we therefore successively have
\begin{eqnarray*}
A & = & \sum_{R\in\mathbf{P}(\L)}
\bbone\{\uz\in R\}\ |\tilde{\rho}(R,\emptyset)|\ 2^{|R|}\\
 & \le & \sum_{R\in\mathbf{P}(\L)}
\bbone\{\uz\in R\}\bbone\{|R|\ge 2\}
2^{\frac{13|R|}{2}-\frac{5}{2}}
\left(J(\zero)+\sqrt{\frac{\l}{2}}\right)^{|R|}\\
 & & \times
\l^{-|R|+\frac{1}{2}}
\times
\sum_{{\mathfrak{T}\leadsto R}\atop{\mathfrak{T}\ {\rm tree}}}
\prod_{\ux\in R} d_{\mathfrak{T}}(\ux)!^{\frac{1}{4}}
\prod_{\{\ux,\uy\}\in\mathfrak{T}} |J(\ux-\uy)|
\end{eqnarray*}
then, introducing the labeling and using Lemma \ref{sumtreedegree},
\[
A\le
\sum_{m\ge 2}  2^{\frac{19}{2}m-\frac{11}{2}}. \l^{-m+\frac{1}{2}}. J_{\neq}^{m-1}
\left(J(\zero)+\sqrt{\frac{\l}{2}}\right)^{m}\ .
\]
For $\l$ large we will have
\[
\frac{2^{\frac{19}{2}} J_{\neq}.\left(J(\zero)+\sqrt{\frac{\l}{2}}\right)}
{\l}
\le\frac{1}{2}
\]
and
\[
||\rho(\cdot,\emptyset)||\le 2^{\frac{29}{2}}  J_{\neq}
\l^{-\frac{3}{2}} \left(J(\zero)+\sqrt{\frac{\l}{2}}\right)^2<1\ .
\]
So we now have,
\begin{eqnarray}
\lefteqn{\mathfrak{W}=
\sum_{\ux_2,\ldots,\ux_n\in\L}
|\<\psi^{\sharp_1}(\zero),\psi^{\sharp_2}(\ux_2),\ldots,\psi^{\sharp_n}(\ux_n)
\>_\L^{\rm T}| } & & \nonumber\\
 & & \le
\sum_{\ux_1,\ldots,\ux_n\in\L}\bbone\{\ux_1=\zero\}
\sum_{p\ge 1}\frac{1}{p!}
\sum_{{R_1,\ldots,R_p\in\mathbf{P}(\L)}\atop{I_1,\ldots,I_p\subset [n]}}\nonumber\\
 & & |\phi^T(R_1,\ldots,R_p)|
\bbone\left\{
\begin{array}{c}
I_q\ {\rm disjoint}\\
\cup I_q=[n]
\end{array}
\right\}
\prod\limits_{q=1}^p\left[
\bbone\left\{
\begin{array}{c}
\forall i\in I_q\\
\ux_i\in R_q
\end{array}
\right\}
\bbone\left\{
\begin{array}{c}
|R_q|\ge 2\\
{\rm or}\ |I_q|\ge 1
\end{array}
\right\}
\right.\nonumber\\
 & & \times
2^{\frac{11}{2}|R_q|+\frac{3}{4}|I_q|-\frac{5}{2}}
\left(J(\zero)+\sqrt{\frac{\l}{2}}\right)^{|R_q|}
\l^{-|R_q|-\frac{|I_q|}{4}+\frac{1}{2}}\nonumber\\
 & & \left. \times
\prod_{\ux\in R_q} c_{I_q}(\ux)!^{\frac{1}{4}}
\times
\sum_{{\mathfrak{T}_q\leadsto R_q}\atop{\mathfrak{T}_q\ {\rm tree}}}
\prod_{\ux\in R_q} d_{\mathfrak{T}_q}(\ux)!^{\frac{1}{4}}
\prod_{\{\ux,\uy\}\in\mathfrak{T}_q} |J(\ux-\uy)|
\right]\ .
\label{monster2}
\end{eqnarray}
Following the same steps as before, including the introduction of $\gamma\in
(0,1]$,
we arrive at
\begin{eqnarray*}
\mathfrak{W} & \le & n!\ .\gamma^{-n}. 2^{\frac{11n}{4}-1}.
\l^{-\frac{n}{4}}
\times\sum_{p\ge 1}\frac{1}{p!}
\sum_{R_1,\ldots,R_p\in \mathbf{P}(\L)}\\
 & &
\bbone\{\zero\in\cup_{q=1}^p R_q\}
|\phi^{\rm T}(R_1,\ldots,R_p)|.
\varsigma(R_1)\ldots\varsigma(R_p)
\end{eqnarray*}
with the nonnegative polymer activities
\begin{eqnarray*}
\varsigma(R) & = & 2^4\gamma.\bbone\{|R|=1\}\times
\l^{-\frac{1}{2}}\left(J(\zero)+\sqrt{\frac{\l}{2}}\right)\\
 & & +\bbone\left\{|R|\ge 2\right\}
2^{\frac{13}{2}|R|-\frac{5}{2}} \left(J(\zero)+\sqrt{\frac{\l}{2}}\right)^{|R|} \\
& & \ \times \l^{-|R|+\frac{1}{2}}
\sum_{{\mathfrak{T}\leadsto R}\atop{\mathfrak{T}\ {\rm tree}}}
\prod_{\ux\in R} d_{\mathfrak{T}}(\ux)!^{\frac{1}{4}}
\prod_{\{\ux,\uy\}\in\mathfrak{T}} |J(\ux-\uy)|\ .
\end{eqnarray*}
Their norm is similarly bounded by
\begin{eqnarray*}
||\varsigma|| & \le & 2^{\frac{9}{2}}\gamma\times
\left(1+J(\zero)\sqrt{\frac{2}{\l}}\right)\\
 & & +\sum_{m\ge 2}
2^{\frac{21}{2}m-\frac{11}{2}}
J_{\neq}^{m-1}\l^{-m+\frac{1}{2}}
\left(J(\zero)+\sqrt{\frac{\l}{2}}\right)^{m}\ .
\end{eqnarray*}
For $\l$ large enough we will have
\[
\frac{2^{\frac{21}{2}} J_{\neq}\left(1+J(\zero)\sqrt{\frac{2}{\l}}\right)}{\l}
\le\frac{1}{2}
\]
and therefore also
\[
||\varsigma||\le
2^{\frac{9}{2}}\gamma\times
\left(1+J(\zero)\sqrt{\frac{2}{\l}}\right)
+2^{\frac{33}{2}}  J_{\neq} \l^{-\frac{3}{2}}
\left(J(\zero)+\sqrt{\frac{\l}{2}}\right)^{2}\ .
\]
The latter expression approaches $2^{\frac{9}{2}} \gamma$
as $\l$ becomes large. Therefore, choosing
$\gamma=\frac{1}{2^{\frac{11}{2}}}$
and applying Lemma \ref{pinsumlemma}
completes the proof of Theorem \ref{largelambdathm}.

\section{The small self-interaction or near-Gaussian regime}

\subsection{The cluster and Mayer expansions for the truncated
correlation functions}
\label{GJSsection}

We will follow a line of argument similar to \S\ref{mayercorrelsec}
in order to obtain a convergent series representation of the truncated
correlations which is adapted to the small $\l$ regime to be considered
in the remainder of this article.
Using the notation of \S\ref{basics} and \S\ref{mayercorrelsec}
we have
$Z_\L(\bm{\alpha})=(\det{\tilde{J}})^{-1} \tilde{Z}_\L(\bm{\alpha})$
where
\[
\tilde{Z}_\L(\bm{\alpha})
=\int_{\mathbb{C}^\L}
 d\mu_C(\psi^*,\psi)
\ e^{-\frac{\l}{4}\sum_{\ux\in\L}|\psi(\ux)|^4}
\prod_{i\in I}\left(1+\alpha_i \psi^{\sharp_i}(\ux_i)\right)\ .
\]
Since $\tilde{J}$ is positive definite, we again have that
${\rm Log}\ \tilde{Z}_\L(\bm{\alpha})$
is analytic in a small polydisc
$D_\L$ around $\bm{\alpha}=\mathbf{0}$ and
\[
\<(\psi^{\sharp_i}(\ux_i))_{i\in I}\>_{\L}^{\rm T}=
\left.\frac{\partial^{|I|}}{\prod_{i\in I} \partial\alpha_i}
\ {\rm Log}\ \tilde{Z}_\L(\bm{\alpha})\right|_{\bm{\alpha}=\mathbf{0}}\ .
\]
The next step is to write an expansion for $\tilde{Z}_\L(\bm{\alpha})$
similar to (\ref{polyrep}). This step is usually called the cluster expansion
in the constructive quantum field theory
literature and was introduced by Glimm,
Jaffe and Spencer~\cite{GJS1,GJS2}.
These expansions which have been simplified and improved over the years
by a small group of experts, are not so well known in the wider
mathematical
community. In what follows we will try to explain the method in
detail, on the
simple case of the lattice field theory considered in
this article. We also use one of the more recent
technical implementations based on the
BKAR
forest interpolation formula of \S\ref{bkarsection}.
This is a kind of combinatorial Taylor expansion with integral
reminder which interpolates between a `complex' fully coupled situation
and a `simpler' fully decoupled one.
In the present case this decoupling expansion
will be applied to the Gaussian measure, since it is the only feature
preventing the random variables of different lattice sites from
being independent.

Before introducing the decoupling expansion for the Gaussian measure, we need
a preliminary expansion of the self-interaction term
$e^{-\frac{\l}{4}\sum_{\ux\in\L}|\psi(\ux)|^4}$.
This is a matter of convenience and spares us the division by the
amplitude of trivial polymers when deriving the polymer gas
representation. This is especially useful in view of the forthcoming
derivatives with respect to the coupling constant $\l$ which one would
rather have act on products instead of ratios.
This preliminary expansion consists in writing
\begin{eqnarray*}
e^{-\frac{\l}{4}\sum_{\ux\in\L}|\psi(\ux)|^4} & = &
\prod_{\ux\in\L}\left[
1+\left(e^{-\frac{\l}{4}|\psi(\ux)|^4}-1
\right)
\right]\\
 & = & \sum_{\U\subset\L}\prod_{\ux\in\U}
\left(e^{-\frac{\l}{4}|\psi(\ux)|^4}-1
\right)\ .
\end{eqnarray*}
Then for each $\ux\in\U$ we write
\[
e^{-\frac{\l}{4}|\psi(\ux)|^4}-1
=\int\limits_0^1\ dt_\ux\ \left(
-\frac{\l}{4}|\psi(\ux)|^4
\right)\
e^{-\frac{\l}{4} t_\ux |\psi(\ux)|^4}\ .
\]
As a result, one has
\[
\tilde{Z}_\L(\bm{\alpha})=
\sum_{\U\subset\L} \left(-\frac{\l}{4}\right)^{|\U|}
\int_{[0,1]^\U}\ d\vec{t}\
\int_{\mathbb{C}^\L}
d\mu_C(\psi^*,\psi)
\]
\begin{equation}
\left(\prod_{i\in I}(1+\alpha_i\psi^{\sharp_i}(\ux_i))
\right)
\left(\prod_{\ux\in\U}
|\psi(\ux)|^4
\right)
\ e^{-\frac{\l}{4}\sum_{\ux\in\U}t_\ux|\psi(\ux)|^4}
\label{prepforbkar}
\end{equation}
where $d\vec{t}$ denotes the Lebesgue measure $\prod_{\ux\in\U} dt_\ux$.

We are now ready to apply Theorem \ref{bkarthm}, using $E=\L$, as follows.
For any multiplet $s=(s_l)_{l\in \L^{(2)}}$ in the closed
convex set $\mathcal{K}_\L$,
we
replace the covariance $C$ by a modified covariance $C[s]$ defined by
\[
C[s](\ux,\ux)=C(\ux,\ux)\ {\rm for\ any}\ \ux\in\L,
\]
and
\[
C[s](\ux,\uy)=s_{\{\ux,\uy\}}C(\ux,\uy)\ {\rm for}\ \ux\neq\uy\ {\rm in\ }\L.
\]
Clearly this new matrix is also Hermitian.
Moreover, the following key positivity property
follows from the previous definitions.

\begin{Lemma}
If $s\in\mathcal{K}_\L$ then $C[s]$ is positive definite.
\end{Lemma}

\noindent{\bf Proof:}
For $s\in\mathcal{K}_\L$ one can find an expression
$s=\sum_{j=1}^k w_j v_{\pi_j} $ with the $w_j$'s nonnegative
and summing up to $1$, and the $\pi_j$ some suitable partitions of $\L$.
Using notations similar to \S\ref{basics}, for any vector
$\psi\in \mathbb{C}^\L$ we have
\begin{eqnarray*}
<\psi,C[s]\psi> & = & \sum_{\ux,\uy\in\L}\psi^*(\ux) C[s](\ux,\uy)\psi(\uy)\\
 & = & \sum_{\ux\in\L}\psi^*(\ux)C(\ux,\ux)\psi(\ux)\\
 & & +\sum_{j=1}^k w_j \sum_{{\ux,\uy\in\L}\atop{\ux\neq\uy}} \psi^*(\ux)
(v_{\pi_j})_{\{\ux,\uy\}}  C(\ux,\uy)\psi(\uy)\\
 & = & \sum_{j=1}^k w_j
\left[
\sum_{\ux\in\L}\psi^*(\ux)C(\ux,\ux)\psi(\ux)\right.\\
 & & \ \left. + \sum_{{\ux,\uy\in\L}\atop{\ux\neq\uy}} \psi^*(\ux)
\left(
\sum_{X\in\pi_j}\bbone\{\ux\in X\}
\bbone\{\uy\in X\}
\right) C(\ux,\uy)\psi(\uy)
\right]\\
 & = & \sum_{j=1}^k w_j \sum_{X\in\pi_j} <\psi_X, C \psi_X>
\end{eqnarray*}
where $\psi_X(\ux)=\bbone\{\ux\in X\} \psi(\ux)$.
Therefore the result is nonnegative since $C$ is positive definite.
Now suppose $<\psi,C[s]\psi>$ vanishes. Since $\sum_{j=1}^k w_j=1$,
one can choose $j_0$ such that $w_{j_0}>0$, and then have for every block
$X\in\pi_{j_0}$ that $<\psi_X, C\psi_X>=0$, i.e., that $\psi_X=0$.
Thus $\psi=\sum_{X\in\pi_{j_0}}\psi_X=0$.
\qed

As a result of this lemma
the corresponding Gaussian measures $d\mu_{C[s]}$ are well
defined.
The last Gaussian integral in (\ref{prepforbkar})
therefore
becomes a function
\[
f(s)=\int_{\mathbb{C}^\L}
d\mu_{C[s]}(\psi^*,\psi)
\left(\prod_{i\in I}(1+\alpha_i\psi^{\sharp_i}(\ux_i))
\right)
\left(\prod_{\ux\in\U}
|\psi(\ux)|^4
\right)
\ e^{-\frac{\l}{4}\sum_{\ux\in\U}t_\ux|\psi(\ux)|^4}
\]
of $s\in\mathcal{K}_\L$
to which one can apply Theorem \ref{bkarthm}.
The well known rule for computing derivatives of Gaussian integrals
with respect to the covariance matrix, namely as a Laplace type
operator acting on the integrand (see e.g.~\cite[\S9.2]{GJ}),
immediately implies the
following representation:
\[
\tilde{Z}_\L(\bm{\alpha})=
\sum_{\U\subset\L} \left(-\frac{\l}{4}\right)^{|\U|}
\int_{[0,1]^\U}\ d\vec{t}\
\sum_{{\F\ {\rm forest}}\atop{{\rm on}\ \L}} \int_{[0,1]^\F}\ d\vec{h}
\int_{\mathbb{C}^\L}
d\mu_{C[s(\F,\vec{h})]}(\psi^*,\psi)
\]
\[
\left(\prod_{l\in\F} \Delta_l
\right)
\left(\prod_{i\in I}(1+\alpha_i\psi^{\sharp_i}(\ux_i))
\right)
\left(\prod_{\ux\in\U}
|\psi(\ux)|^4
\right)
\ e^{-\frac{\l}{4}\sum_{\ux\in\U}t_\ux|\psi(\ux)|^4}\ .
\]
Here we denoted by $\Delta_l$ the operator of Laplace type given,
for any unordered pair of distinct sites $\ux$ and $\uy$ by
\begin{equation}
\Delta_{\{\ux,\uy\}}=
C(\ux,\uy)\frac{\partial}{\partial\psi(\ux)}
\frac{\partial}{\partial\psi^*(\uy)}
+C(\uy,\ux)\frac{\partial}{\partial\psi(\uy)}
\frac{\partial}{\partial\psi^*(\ux)}
\label{deltasplit}
\end{equation}
using the standard $\frac{\partial}{\partial z}$ and
$\frac{\partial}{\partial\bar{z}}$
vector fields of multivariate complex analysis.
The differential operators act on everything to their right.

Now, let $\pi$ be the partition of $\L$
into connected components of the forest $\F$. An important feature of
the
definition of $s(\F,\vec{h})$ is that it vanishes between components.
This implies the componentwise factorization of the Gaussian integral.
One can also factorize the different combinatorial sums involved,
i.e.,
those over the sets $\U_R=\U\cap R$, as well as the
ones
over the trees $\T_R$ connecting each $R\in\pi$ and
which together make up the
forest $\F$.
In sum, one has
\begin{equation}
\tilde{Z}_\L(\bm{\alpha})=\sum_{\pi\in\Pi_\L}
\prod_{R\in\pi} \zeta_0(R,\bm{\alpha},\l)
\label{ztildepartition}
\end{equation}
where for each polymer $R$
we defined the polymer activity
\[
\zeta_0(R,\bm{\alpha},\l)=
\sum_{\U\subset R}
\sum_{J\subset I_R}
\sum_{{\T\leadsto R}\atop{\T\ {\rm tree}}}
\left(-\frac{\l}{4}\right)^{|\U|}
\left(\prod_{i\in J}\alpha_i\right)
\int_{[0,1]^\U}\ d\vec{t}\
\int_{[0,1]^\T}\ d\vec{h}
\]
\begin{equation}
\int_{\mathbb{C}^R}
d\mu_{C[s(\T,\vec{h})]}
(\psi^*,\psi)\
\left(\prod_{l\in\T} \Delta_l
\right)
\left(\prod_{i\in J}\psi^{\sharp_i}(\ux_i)
\right)
\left(\prod_{\ux\in\U}
|\psi(\ux)|^4
\right)
\ e^{-\frac{\l}{4}\sum_{\ux\in\U}t_\ux|\psi(\ux)|^4}
\label{zetazero}
\end{equation}
again with $I_R=\{i\in I| \ux_i\in R\}$.
Note that the covariance $C$ which is used here is the restriction
to $R$ of the one defined on $\L$ as the inverse of $\tilde{J}$. It
therefore retains a slight dependence on the volume $\L$ which
contains the polymer
$R$. We nevertheless suppressed it in the notation for better
readability.

We now introduce in (\ref{zetazero}) the decomposition
\[
1=
\bbone\left\{
\begin{array}{c}
|R|=1\ {\rm and}\ \\
\U=\emptyset\ {\rm and}\ \\
J=\emptyset
\end{array}
\right\}
+
\bbone\left\{
\begin{array}{c}
|R|\ge 2\ {\rm or}\ \\
|\U|\ge 1\ {\rm or}\ \\
|J|\ge 1
\end{array}
\right\}
\]
and break $\zeta_0(R,\bm{\alpha},\l)$
accordingly as a sum of two contributions.
It is easy to see, because all sums and integrals become trivial and also
because even for a single site the Gaussian measures are normalized,
that the first contribution reduces to $\bbone\{|R|=1\}$.
Therefore, by only keeping track of polymers for which the second
contribution is selected,
one can rewrite (\ref{ztildepartition}) as
a polymer gas representation similar to (\ref{polyrep}), namely
\begin{equation}
\tilde{Z}_{\L}(\bm{\alpha})=
\sum_{p\ge 0}\frac{1}{p!}
\sum_{R_1,\ldots,R_p\in\mathbf{P}(\L)}
\bbone\left\{
\begin{array}{c}
{\rm the}\ R_i\ {\rm are}\\
{\rm disjoint}
\end{array}
\right\}
\zeta(R_1,\bm{\alpha},\l)\ldots
\zeta(R_p,\bm{\alpha},\l)
\label{cqftcluster}
\end{equation}
with
the polymer activities defined as
\begin{eqnarray}
\lefteqn{
\zeta(R,\bm{\alpha},\l)=
\sum_{\U\subset R}
\sum_{J\subset I_R}
\sum_{{\T\leadsto R}\atop{\T\ {\rm tree}}}
\bbone\left\{
\begin{array}{c}
|R|\ge 2\ {\rm or}\ \\
|\U|\ge 1\ {\rm or}\ \\
|J|\ge 1
\end{array}
\right\}} & & \nonumber\\
 & & \times
\left(-\frac{\l}{4}\right)^{|\U|}
\left(\prod_{i\in J}\alpha_i\right)
\int_{[0,1]^\U}\ d\vec{t}\
\int_{[0,1]^\T}\ d\vec{h}
\int_{\mathbb{C}^R}
d\mu_{C[s(\T,\vec{h})]}
(\psi^*,\psi) \nonumber\\
& & \left(\prod_{l\in\T} \Delta_l
\right)
\left(\prod_{i\in J}\psi^{\sharp_i}(\ux_i)
\right)
\left(\prod_{\ux\in\U}
|\psi(\ux)|^4
\right)
\ e^{-\frac{\l}{4}\sum_{\ux\in\U}t_\ux|\psi(\ux)|^4}\ .
\end{eqnarray}
Now the same line of argument leading up to Proposition \ref{correlseries}
shows the following.

\begin{Proposition}
\label{cqftcorrseries}
Define for any subset $J$ of the source label set $I$, and any polymer
$R\in\mathbf{P}(\L)$,
\begin{eqnarray}
\lefteqn{
\tilde{\zeta}(R,J,\l)=
\bbone\left\{
\begin{array}{c}
\forall i\in J\\
\ux_i\in R
\end{array}
\right\}
\sum_{\U\subset R}
\left(-\frac{\l}{4}\right)^{|\U|}
\bbone\left\{
\begin{array}{c}
|R|\ge 2\ {\rm or}\ \\
|\U|\ge 1\ {\rm or}\ \\
|J|\ge 1
\end{array}
\right\}} & & \nonumber\\
 & & \times \sum_{{\T\leadsto R}\atop{\T\ {\rm tree}}}
\int_{[0,1]^\U}\ d\vec{t}\
\int_{[0,1]^\T}\ d\vec{h}
\int_{\mathbb{C}^R}
d\mu_{C[s(\T,\vec{h})]}
(\psi^*,\psi) \nonumber\\
 & & \left(\prod_{l\in\T} \Delta_l
\right)
\left(\prod_{i\in J}\psi^{\sharp_i}(\ux_i)
\right)
\left(\prod_{\ux\in\U}
|\psi(\ux)|^4
\right)
\ e^{-\frac{\l}{4}\sum_{\ux\in\U}t_\ux|\psi(\ux)|^4}\ .
\label{zetatildedef}
\end{eqnarray}
Provided the source-free condition
\begin{equation}
\sup_{\ux\in\L}\sum_{R\in\mathbf{P}(\L)}
\bbone\{\ux\in R\}\ |\tilde{\zeta}(R,\emptyset,\l)|\ 2^{|R|}<1
\label{cvcriterionzeta}
\end{equation}
holds, one has the absolutely convergent series representation
for all truncated correlation functions
\begin{eqnarray}
\lefteqn{
\<(\psi^{\sharp_i}(\ux_i))_{i\in I}\>_{\L}^{\rm T}=
\sum_{p\ge 1}\frac{1}{p!}
\sum_{{R_1,\ldots,R_p\in\mathbf{P}(\L)}\atop{I_1,\ldots,I_p\subset I}}
\phi^{\rm T}(R_1,\ldots,R_p)} & & \nonumber\\
 & & \times \bbone\left\{
\begin{array}{c}
I_q\ {\rm disjoint}\\
\cup I_q=I
\end{array}
\right\}
\prod_{q=1}^p \tilde{\zeta}(R_q,I_q,\l)\ .
\label{cqftmayer}
\end{eqnarray}
\end{Proposition}

\begin{Remark}
Although this is not really necessary for the proof of Proposition
\ref{cqftcorrseries}, note that the source-free activities
$\tilde{\zeta}(R,\emptyset,\l)$ are real-valued. This is because
of the hypothesis $J(-\ux)=J(\ux)^*$ which implies
that the differential operators $\Delta_{\{\ux,\uy\}}$ preserve
the real-valuedness of functions.
\end{Remark}

\begin{Remark}
In the constructive quantum field theory literature, an equation such
as (\ref{cqftcluster}) would be called a cluster expansion for the partition
function $\tilde{Z}(\bm{\alpha})$ which is written as a sum over collections
of disjoint polymers.  An equation such as (\ref{cqftmayer}) involving
sums over collections of polymers with the coefficient $\phi^{\rm T}$
would be called a Mayer expansion for the truncated correlation functions.
\end{Remark}

We now go back to the setting of Theorem \ref{smalllambdathm} and also
restore the $\l$ dependence in the notation for correlation functions.
In order to extract the factor $\l^N$
in the clustering estimate, we will perform an additional
Taylor expansion of the connected correlation function.
We write
\[
\<\psi^{\sharp_1}(\ux_1),\ldots,\psi^{\sharp_n}(\ux_n)\>_{\L,\l}^{\rm T}=
\sum\limits_{k=0}^{N-1}
\frac{\l^k}{k!}
\left(\frac{d}{d\l}
\right)^k
\left.
\<\psi^{\sharp_1}(\ux_1),\ldots,\psi^{\sharp_n}(\ux_n)\>_{\L,\l}^{\rm T}
\right|_{\l=0}
\]
\[
+\int\limits_0^1 \ du\ \frac{(1-u)^{N-1}}{(N-1)!}
\left(\frac{d}{du}
\right)^N
\<\psi^{\sharp_1}(\ux_1),\ldots,\psi^{\sharp_n}(\ux_n)\>_{\L,u\l}^{\rm T}\ .
\]
However, by Lemma \ref{Benedettohmwk} and the hypothesis $n\ge 2(N+1)$,
the Taylor polynomial vanishes and the right-hand side reduces to the
integral remainder, i.e.,
\begin{eqnarray*}
\lefteqn{
\<\psi^{\sharp_1}(\ux_1),\ldots,
\psi^{\sharp_n}(\ux_n)\>_{\L,\l}^{\rm T}=} & & \nonumber\\
 & & \int\limits_0^1 \ du\ \frac{(1-u)^{N-1}}{(N-1)!}
\left(\frac{d}{du}
\right)^N
\<\psi^{\sharp_1}(\ux_1),\ldots,\psi^{\sharp_n}(\ux_n)\>_{\L,u\l}^{\rm T}\ .
\end{eqnarray*}
So far we only used the fact that correlations as $C^\infty$ functions
of $\l$ on $[0,+\infty)$. However, the next step is to use the representation
(\ref{cqftmayer}), and also to differentiate it, term by term, $N$ times.
This will require some estimates, in order
to justify the convergence criterion
(\ref{cvcriterionzeta}) as well as the following outcome of term by term
differentiation
\begin{eqnarray}
\lefteqn{
\<\psi^{\sharp_1}(\ux_1),\ldots,
\psi^{\sharp_n}(\ux_n)\>_{\L,\l}^{\rm T}=} & & \nonumber\\
 & & \int\limits_0^1 \ du\ \frac{(1-u)^{N-1}}{(N-1)!}
\sum_{p\ge 1}\frac{1}{p!}
\sum_{{R_1,\ldots,R_p\in\mathbf{P}(\L)}\atop{I_1,\ldots,I_p\subset[n]}}
\phi^{\rm T}(R_1,\ldots,R_p) \nonumber\\
 & & \times\bbone
\left\{
\begin{array}{c}
I_q\ {\rm disjoint}\\
\cup I_q=[n]
\end{array}
\right\}
\sum_{N_1+\cdots+N_p=N}\frac{N!}{N_1!\ldots N_p!}
\prod_{q=1}^p
\left(\frac{d}{du}\right)^{N_q}
\tilde{\zeta}(R_q,I_q,u\l)
\label{termbyterm}
\end{eqnarray}
for $n\ge 2(N+1)$.
Finally we will use this last representation as input for the $l^1$-clustering
property.

\subsection{The estimates}
This section will provide the necessary estimates for the proof of
Theorem \ref{smalllambdathm}.
Many of the ideas used in these estimates originated in the work
Glimm-Jaffe-Spencer (see e.g.~\cite[Ch. 18]{GJ},
\cite[\S III.1]{Rivasseaubook}, and~\cite{AR2}).
We will first suppose that the $J$ function governing the Gaussian
measure is so chosen that the
constant $K_0$ appearing in Lemma \ref{decaylemma} is equal to $1$.
Later in \S\ref{finalclusterbound} we will get rid of this
restriction by a simple scaling transformation on the field variable $\psi$.

\subsubsection{The bound on a single polymer activity and its derivatives}
The basic quantity we need to bound is
$\left(\frac{d}{du}\right)^{M}
\tilde{\zeta}(R,I,u\l)$,
for any integer $M\ge 0$, polymer $R$, index set
$I\in[n]$, and $u\in [0,1]$.
From (\ref{zetatildedef}) one easily gets
\[
\left(\frac{d}{du}\right)^{M}
\tilde{\zeta}(R,I,u\l)=
\bbone\left\{
\begin{array}{c}
\forall i\in I\\
\ux_i\in R
\end{array}
\right\}
\sum_{\U\subset R}
\bbone\left\{
\begin{array}{c}
|R|\ge 2\ {\rm or}\ \\
|\U|\ge 1\ {\rm or}\ \\
|I|\ge 1
\end{array}
\right\}
\]
\[
\sum\limits_{k=0}^{M}\frac{M!}{k!(M-k)!}
\bbone\{|\U|\ge M-k\}
\frac{|\U|!}{(|\U|-M+k)!} u^{|\U|-M+k}
\left(-\frac{\l}{4}\right)^{|\U|+k}
\]
\begin{equation}
\sum_{{\T\leadsto R}\atop{\T\ {\rm tree}}}
\sum_{\uy_1,\ldots,\uy_k\in\U}
\int_{[0,1]^\U}\ d\vec{t}\
\int_{[0,1]^\T}\ d\vec{h}
\left(\prod_{j=1}^{k}
t_{\uy_j}
\right)\ \cI
\label{derrho}
\end{equation}
where $\cI$ refers to the remaining Gaussian integral, namely,
\[
\cI=\int_{\mathbb{C}^R}
d\mu_{C[s(\T,\vec{h})]}
(\psi^*,\psi)\
\left(\prod_{l\in\T} \Delta_l
\right)
\left(\prod_{i\in I}\psi^{\sharp_i}(\ux_i)
\right)
\]
\begin{equation}
\left(\prod_{\ux\in\U}
|\psi(\ux)|^4
\right)
\left(\prod_{j=1}^{k}
|\psi(\uy_j)|^4
\right)
\ e^{-\frac{\l u}{4}\sum_{\ux\in\U}t_\ux|\psi(\ux)|^4}\ .
\label{Iintegral}
\end{equation}
When one performs the derivatives coming from the $\Delta_l$ operators,
this further splits into
\[
\cI=\sum_{\gp}\cI_\gp
\]
where, avoiding excessive formalization, we denoted by $\gp$
any derivation procedure including the detailed information as to which
specific $\psi$ or $\psi^*$ factor has been destroyed, and by which
$\frac{\partial}{\partial\psi}$
or $\frac{\partial}{\partial\psi^*}$ operator.
We will use a bound of the form
\[
|\cI|\le \left(\max_{\gp} |\cI_\gp|\right)
\sum_{\gp} 1\ .
\]
Note that a term $\cI_\gp$ has the form
\begin{eqnarray*}
\lefteqn{
\cI_\gp=C_\gp . L_\gp .
\int_{\mathbb{C}^R}
d\mu_{C[s(\T,\vec{h})]}
(\psi^*,\psi)} & & \\
 & & \left(\prod_{\ux\in R}\{\psi(\ux)^{m(\ux)}\psi^*(\ux)^{m^*(\ux)}\}\right)
\ e^{-\frac{\l u}{4}\sum_{\ux\in\U}t_\ux|\psi(\ux)|^4}
\end{eqnarray*}
where the $m(\ux)$ and $m^*(\ux)$ are some local multiplicities and
$L_\gp$ is a product of factors $-\frac{\l u t_\ux}{4}$ produced by the
derivatives
which acted on the exponential.
Finally, $C_\gp$ is a product of propagators corresponding
to the edges in the tree $\T$. It depends on whether,
for each $\Delta_l$, one choses the $C(\ux,\uy)$ or the $C(\uy,\ux)$ term.
In any case, Lemma \ref{decaylemma} together with the assumption $K_0=1$
implies
\[
|C_\gp|\le
e^{-\mu_0\sum_{\{\ux,\uy\}\in\T} |\ux-\uy|}\ .
\]
We will use the notation $|m|=\sum_{\ux\in R}m(\ux)$ for the total
multiplicity
of a multiindex such as $(m(\ux))_{\ux\in R}$.
Using the positivity of the interaction, and in particular
the positivity of $\l$, one has
\[
0<
\ e^{-\frac{\l u}{4}\sum_{\ux\in\U}t_\ux|\psi(\ux)|^4}
\le 1\ .
\]
Therefore, by the Cauchy-Schwartz inequality
\[
\left|
\int_{\mathbb{C}^R}
d\mu_{C[s(\T,\vec{h})]}
(\psi^*,\psi)\
\prod_{\ux\in R}\{
\psi(\ux)^{m(\ux)}\psi^*(\ux)^{m^*(\ux)}
\}
\ e^{-\frac{\l u}{4}\sum_{\ux\in\U}t_\ux|\psi(\ux)|^4}
\right|
\]
\[
\le
\left[
\int_{\mathbb{C}^R}
d\mu_{C[s(\T,\vec{h})]}
(\psi^*,\psi)\
\prod_{\ux\in R}\{
\psi(\ux)^{\hat{m}(\ux)}\psi^*(\ux)^{\hat{m}(\ux)}
\}
\right]^{\frac{1}{2}}
\]
where $\hat{m}(\ux)=m(\ux)+m^*(\ux)$.

From now on we also impose the hypothesis $0<\l\le 4$, which implies
$|L_\gp|\le 1$.
We now invoke the following classical lemma of constructive field
theory~\cite{GJS2,EMS} also called the principle of local factorials,
in order to bound the last Gaussian integral.
\begin{Lemma}\label{locfact}
For any $q$, and collection of sites $\uz_1,\ldots,\uz_q$ and
$\uw_1,\ldots,\uw_q$ in $R$,
one has
\[
\left|
\int_{\mathbb{C}^R}
d\mu_{C[s(\T,\vec{h})]}
(\psi^*,\psi)\
\psi(\uz_1)\ldots\psi(\uz_q)
\psi^*(\uw_1)\ldots\psi^*(\uw_q)
\right|\le
K_1^q\prod_{\ux\in R} n^*(\ux)!
\]
where $n^*(\ux)$
denotes the number of indices $i$, $1\le i\le q$, such that
$\uw_i=\ux$
and
$K_1=K_1(d,\mu_0)$ of (\ref{expdecayeq}).
\end{Lemma}

\noindent{\bf Proof:}
By the Isserlis-Wick Theorem we have
\[
\left|\int_{\mathbb{C}^R}
d\mu_{C[s(\T,\vec{h})]}(\psi^*,\psi)\
\psi(\uz_1)\ldots\psi(\uz_q)
\psi^*(\uw_1)\ldots\psi^*(\uw_q)\right|\qquad
\]
\[
\qquad\le \sum_{\sigma\in\mathfrak{S}_q}
\prod_{i=1}^q |C(\uz_i,\uw_{\sigma(i)})|
\le \sum_{\sigma\in\mathfrak{S}_q}
e^{-\mu_0 \sum_{i=1}^q |\uz_i-\uw_{\sigma(i)}|}
\]
because the decoupling parameters in $s(\T,\vec{h})$ are between $0$
and $1$, and also because of Lemma \ref{decaylemma}.
Now
\[
\sum_{\sigma\in\mathfrak{S}_q}
e^{-\mu_0 \sum_{i=1}^q |\uz_i-\uw_{\sigma(i)}|}=
\sum_{\uu_1,\ldots,\uu_q\in R}
e^{-\mu_0 \sum_{i=1}^q |\uz_i-\uu_i|}
\sum_{\sigma\in\mathfrak{S}_q}
\bbone\left\{
\forall i, \uw_{\sigma(i)}=\uu_i
\right\}
\]
and the last sum over $\sigma$ is either equal to
$\prod_{\ux\in R} n^*(\ux)!$ or vanishes,
depending on whether or not the $\uu$ sequence is a permutation of the
$\uw$ sequence.
Now the claim follows using (\ref{expdecayeq}).
\qed

We now apply Lemma \ref{locfact} to get the following bound
\[
|\cI_\gp|\le
e^{-\mu_0\sum_{\{\ux,\uy\}\in\T} |\ux-\uy|}
\times
K_1^{|m|}\times
\prod_{\ux\in R}\left(
m(\ux)+m^*(\ux)
\right)!^{\frac{1}{2}}
\]
where we used the easily verifyable fact
$|m|=|m^*|=\frac{|\hat{m}|}{2}$, i.e., in any
$\cI_\gp$ one always has an equal number of $\psi$ and $\psi^*$
factors remaining.
Let us define the following initial multiplicities, i.e., before
applying the derivatives, for the $\psi$ and $\psi^*$ fields respectively:
\[
m_0(\ux)=|\{i\in I|\ux_i=\ux\ {\rm and}\ \sharp_i=\emptyset\}|+
2\bbone_{\{\ux\in\U\}}+2|\{j,1\le j\le k|\uy_j=\ux\}|
\]
and
\[
m_0^*(\ux)=|\{i\in I|\ux_i=\ux\ {\rm and}\ \sharp_i=*\}|+
2\bbone_{\{\ux\in\U\}}+2|\{j,1\le j\le k|\uy_j=\ux\}|\ .
\]
Once again we have $|m_0|=|m_0^*|$.
When expanding the sums (\ref{deltasplit}), each site will receive
a number $\de(\ux)$ of $\frac{\partial}{\partial\psi(\ux)}$
derivatives, as well as a number $\de^*(\ux)$ of
$\frac{\partial}{\partial\psi^*(\ux)}$ derivatives.
Besides, one trivially has
\[
\de(\ux)+\de^*(\ux)=d(\ux)
\]
where $d(\ux)$ denotes the degree of the vertex $\ux\in R$
in the tree $\T$.

Now it is easy to see that the maximal number of field factors
occurs when all the derivatives pull new $|\psi|^4$ vertices from
the exponential. Namely, for any $\ux\in R$,
\begin{eqnarray*}
m(\ux) & \le & m_0(\ux)+2\de^*(\ux)+\de(\ux)\ ,\\
m^*(\ux) & \le & m_0^*(\ux)+\de^*(\ux)+2\de(\ux)\ ,\\
m(\ux)+m^*(\ux) & \le & m_0(\ux)+m_0^*(\ux)+3d(\ux)\ .
\end{eqnarray*}

We will also need the identities
\begin{equation}
\sum_{\ux\in R}
(m_0(\ux)+m_0^*(\ux))=
|I|+4|\U|+4k
\label{summzero}
\end{equation}
and
\begin{equation}
\sum_{\ux\in R}d(\ux)=2(|R|-1)
\label{sumd}
\end{equation}
which imply
\begin{equation}
|m_0|+|m_0^*|+3|d|=
|I|+4|\U|+4k+6|R|-6
\label{mzerosandd}
\end{equation}
and
\[
|m|\le\frac{|I|}{2}+2|\U|+2k+3|R|-3\ .
\]

We are now ready to bound the number of derivation procedures $\gp$.

\begin{Lemma}
The number of derivation procedures is bounded by
\[
\sum_\gp 1\le
2^{|R|-1}e^{|I|+4|\U|+4k+6|R|-3}
\prod_{\ux\in R} d(\ux)!\ .
\]
\end{Lemma}

\noindent{\bf Proof:}
One pays a factor $2^{|R|-1}$ for the sums (\ref{deltasplit}).
For $\ux\in R\backslash \U$ the derivatives can only act on the fields
already present, and we therefore have a number of choices limited by
\[
\bbone_{\{m_0(\ux)\ge\de(\ux)\}}\bbone_{\{m_0^*(\ux)\ge\de^*(\ux)\}}
\frac{m_0(\ux)!\ m_0^*(\ux)!}
{(m_0(\ux)-\de(\ux))!\ (m_0^*(\ux)-\de^*(\ux))!}\ .
\]
For $\ux\in\U$,
the number of terms produced by applying the derivatives is bounded by
\begin{equation}
d(\ux)!\ \times\ \exp\left\{
m_0(\ux)+m_0^*(\ux)+\frac{3}{2}d(\ux)+3
\right\}\ .
\label{derbdpersite}
\end{equation}
Indeed, this is the number of terms produced when computing
\[
\left(
\frac{\partial}{\partial\psi(\ux)}
\right)^{\de(\ux)}
\left(
\frac{\partial}{\partial\psi^*(\ux)}
\right)^{\de^*(\ux)}
\psi(\ux)^{m_0(\ux)}\psi^*(\ux)^{m_0^*(\ux)}\
e^{-\frac{\l u}{4} \psi(\ux)^2\psi^*(\ux)^2}\ .
\]
Let us first perform the $\frac{\partial}{\partial\psi^*(\ux)}$
derivatives and then the $\frac{\partial}{\partial\psi(\ux)}$
derivatives.
When evaluating the very last derivative
$\frac{\partial}{\partial\psi(\ux)}$
one has to choose between deriving a new vertex from the exponential
which gives a factor $2$ for the choice of $\psi$ in the
$\psi(\ux)^2\psi^*(\ux)^2$ vertex, or deriving a field factor which
was already there which at most gives
$m_0(\ux)+2\de^*(\ux)+\de(\ux)-1$
possibilities. Indeed, either the factor was present initially
which corresponds to $m_0(\ux)$ choices, or it was in a vertex first
derived from the exponential by a
$\frac{\partial}{\partial\psi^*(\ux)}$
derivative, in which case one has to pay a factor $\de^*(\ux)$
to identify that derivative and a factor $2$ for the choice of field $\psi$
within the vertex $-\frac{\l u}{4} \psi(\ux)^2\psi^*(\ux)^2$. The last
possibility
is when the derived factor was in a vertex first produced by one of
the previous $\de(\ux)-1$ derivatives
of type $\frac{\partial}{\partial\psi(\ux)}$.
Such a derivative already consumes one of the two $\psi$'s in the
vertex, so we only have to pay a factor of $\de(\ux)-1$.
In sum, the last $\frac{\partial}{\partial\psi(\ux)}$ derivative at
most gives $(m_0(\ux)+2\de^*(\ux)+\de(\ux)+1)$ possibilities.
Likewise the before last $\frac{\partial}{\partial\psi(\ux)}$
derivatives has at most $(m_0(\ux)+2\de^*(\ux)+\de(\ux))$ options, etc.
Therefore, the number of possibilities for the
$\frac{\partial}{\partial\psi(\ux)}$ derivatives is bounded by
$\frac{(m_0(\ux)+2\de^*(\ux)+\de(\ux)+1)!}{(m_0(\ux)+2\de^*(\ux)+1)!}$.
By a similar reasoning, that of the
$\frac{\partial}{\partial\psi^*(\ux)}$
derivatives, which are performed first, is bounded by
$\frac{(m_0^*(\ux)+\de^*(\ux)+1)!}{(m_0^*(\ux)+1)!}$.
As a result, the total number of possibilities is at most
\[
\frac{(m_0(\ux)+2\de^*(\ux)+\de(\ux)+1)!}{(m_0(\ux)+2\de^*(\ux)+1)!}
\times
\frac{(m_0^*(\ux)+\de^*(\ux)+1)!}{(m_0^*(\ux)+1)!}
\]
\begin{equation}
\le
\left[m_0(\ux)+2\de^*(\ux)+\frac{\de(\ux)}{2}+\frac{3}{2}
\right]^{\de(\ux)}
\left[m_0^*(\ux)+\frac{\de^*(\ux)}{2}+\frac{3}{2}
\right]^{\de^*(\ux)}
\label{locderbound}
\end{equation}
where we used the arithmetic versus geometric mean inequality
\begin{eqnarray*}
\frac{s!}{(s-q)!} & = & s(s-1)\cdots(s-q+1)\\
 & \le & \left[\frac{s+(s-1)+\cdots+(s-q+1)}{q}\right]^q\\
 & \le & \left(s-\frac{q-1}{2}\right)^q\ .
\end{eqnarray*}
Finally using the inequality $x^n\le n!e^x$ for each of the two factors
on the right hand side of (\ref{locderbound}),
as well as the trivial inequality
$\de(\ux)!\de^*(\ux)!\le d(\ux)!$, we find that the number of terms
produced
by the derivatives at the site $\ux$ is bounded by
\[
d(\ux)!\ \times\ \exp\left\{
m_0(\ux)+m_0^*(\ux)+\frac{5}{2}\de^*(\ux)+\frac{1}{2}\de(\ux)+3
\right\}\ .
\]
Now redo the same reasoning, but this time first applying the
$\frac{\partial}{\partial\psi(\ux)}$ and then the
$\frac{\partial}{\partial\psi^*(\ux)}$ derivatives. One will get the
same bound but with $\de(\ux)$ and $\de^*(\ux)$ exchanged.
Taking the geometric mean of the two bounds gives the desired
estimate (\ref{derbdpersite}). Using the same inequalities, the bound
for $\ux\in R\backslash \U$ is easily seen to be no greater than the
one
for the $\ux\in\U$ case.
The lemma now follows from
(\ref{summzero}) and (\ref{sumd}).
\qed

Putting the previous considerations together we now have a bound on $\cI$
from (\ref{Iintegral}):
\[
|\cI|\le
2^{|R|-1}
e^{|I|+4|\U|+4k+6|R|-3}
K_1^{\frac{1}{2}|I|+2|\U|+2k+3|R|-3}
\]
\[
\times
e^{-\mu_0\sum_{\{\ux,\uy\}\in\T} |\ux-\uy|}
\times
\prod_{\ux\in R} d(\ux)!
\times
\prod_{\ux\in R}
\left(m_0(\ux)+m_0^*(\ux)+3d(\ux)
\right)!^{\frac{1}{2}}\ .
\]
Note that we used $K_1\ge 1$ which is clear
from (\ref{expdecayeq}) since $\sum_{\uz\in\Z^d} e^{-\mu|\uz|}\ge 1$.
We now use the estimate
\[
\left(m_0(\ux)+m_0^*(\ux)+3d(\ux)
\right)!\le \left(m_0(\ux)+m_0^*(\ux)\right)! \times
d(\ux)!^3\times
4^{m_0(\ux)+m_0^*(\ux)+3d(\ux)}
\]
and identity (\ref{mzerosandd})
in order to write
\begin{eqnarray*}
\lefteqn{
\prod_{\ux\in R}
\left(m_0(\ux)+m_0^*(\ux)+3d(\ux)
\right)!^{\frac{1}{2}}\le
2^{|I|+4|\U|+4k+6|R|-6}} & & \\
 & & \times \prod_{\ux\in R}
d(\ux)!^{\frac{3}{2}}\times
\prod_{\ux\in R}
\left(m_0(\ux)+m_0^*(\ux)\right)!^{\frac{1}{2}}\ .
\end{eqnarray*}
We also need the following important remark: if $|R|\ge 2$, then the integral
$\mathcal{I}$ of (\ref{Iintegral}) is zero unless $|R|\le |\U|+|I|+k$.
Indeed, if $|R|\ge 2$ then the connecting tree $\T$ is nonempty and therefore
each site $\ux\in R$ receives at least one field derivative.
If such a site $\ux$ contains no source, i.e., $\ux\neq\ux_i$, for
any $i\in I$, and if the site is not in $\U$, and if $\ux\neq\uy_j$ for any
$j$, $1\le j\le k$,
then the derivative has
nothing to act on and the integral $\cI$ is zero.
Note that if $|R|=1$, then we already have a characteristic function
in (\ref{derrho})
enforcing $|\U|\ge 1$ or $|I|\ge 1$. As a result, we always have
$|R|\le |\U|+|I|+k$, which can be exploited by introducing the corresponding
characteristic function.
This condition, together with $|\U|\ge M-k$ and the assumption $0<\l\le 4$
implies
\begin{equation}
\left(\frac{\l}{4}\right)^{|\U|+k}\le
\left(\frac{\l}{4}\right)^{\max\{|R|-|I|,M\}}\ .
\label{smallfactors}
\end{equation}
The previous considerations allow us to write
\[
\left|
\left(\frac{d}{du}\right)^{M}
\tilde{\zeta}(R,I, u\l)
\right|\le
\bbone\left\{
\begin{array}{c}
\forall i\in I\\
\ux_i\in R
\end{array}
\right\}
\sum_{\U\subset R}
\bbone\left\{
\begin{array}{c}
|R|\ge 2\ {\rm or}\ \\
|\U|\ge 1\ {\rm or}\ \\
|I|\ge 1
\end{array}
\right\}
\]
\[
\sum\limits_{k=0}^{M}
\bbone\{|\U|\ge M-k\}\ \frac{M!}{k!}\ 2^{|\U|}
\ \left(\frac{\l}{4}\right)^{\max\{|R|-|I|,M\}}
\ \bbone\{|R|\le |\U|+|I|+k\}
\]
\[
\sum_{{\T\leadsto R}\atop{\T\ {\rm tree}}}
\sum_{\uy_1,\ldots,\uy_k\in\U}
2^{|R|-1}\times
(2e\sqrt{K_1})^{|I|+4|\U|+4k+6|R|-3}\times
e^{-\mu_0\sum_{\{\ux,\uy\}\in\T} |\ux-\uy|}
\]
\begin{equation}
\times
\prod_{\ux\in R}
d(\ux)!^{\frac{5}{2}}\times
\prod_{\ux\in R}
\left(m_0(\ux)+m_0^*(\ux)
\right)!^{\frac{1}{2}}
\label{rhobound}
\end{equation}
where we used $K_1\ge 1$ as well as the inequality
$\frac{|\U|!}{(M-k)!(|\U|-M+k)!}\le 2^{|\U|}$.

We now need a lemma which bounds local factorials of the degrees in
the tree by a portion of the tree decay. This is a volume effect due
to the finite dimensionality of the host lattice $\Z^d$.

\begin{Lemma}\label{volumeeffect}
For any $\alpha>0$ we have
\[
\prod_{\ux\in R}
d(\ux)!
\ \times
e^{-\alpha\sum_{\{\ux,\uy\}\in\T} |\ux-\uy|}
\le K_2^{|R|}
\]
where $K_2=\max\{K_{2,1},K_{2,2}\}$
with
\[
K_{2,1}=
\left(
\left\lfloor
\frac{2\pi^{\frac{d}{2}}d^{\frac{d}{2}}}{\Gamma\left(\frac{d}{2}+1\right)}
\right\rfloor
\right)!
\]
and
\[
K_{2,2}=\exp\left(
\sup_{x\in[1,+\infty[}\left\{
x\log x-
\frac{\alpha \Gamma\left(\frac{d}{2}+1\right)^{\frac{1}{d}}}
{2^{2+\frac{1}{d}}\sqrt{\pi}}
x^{1+\frac{1}{d}}+\frac{\alpha\sqrt{d}}{4}x
\right\}
\right)\ .
\]
\end{Lemma}

\noindent{\bf Proof:}
We write the quantity to be estimated as the product over $\ux\in R$
of
\begin{equation}
d(\ux)!\ e^{-\frac{\alpha}{2}\sum_{\uy|\{\ux,\uy\}\in\T}|\ux-\uy|}
\label{xandfriends}
\end{equation}
and we will bound the last expression using the fact that the $d(\ux)$
sites $\uy$ which are neighbors of $\ux$ in the tree $\T$ are
distinct,
and the more they are the further away from $\ux$ they have to be.
Indeed, for any $r\ge 0$, the number $B_r$ of lattice points
at distance at most $r$ from $\ux$ satisfies
\begin{eqnarray*}
B_r & \le & 2^d\left|
\left\{\uz\in\mathbb{N}^d, |\uz|\le r
\right\}\right|\\
 & \le & 2^d\ {\rm Vol}\left(
\bigcup\limits_{\uz\in\mathbb{N}^d, |\uz|\le r}
\ \uz+[0,1[^d\right)\\
 & \le & 2^d\ {\rm Vol}\left(
\left\{\uz\in\mathbb{R}_+^d, |\uz|\le r+\sqrt{d}\right\}
\right)\\
 & \le & {\rm Vol}(S^{d-1})\int\limits_{0}^{r+\sqrt{d}}
\ d\rho\ \rho^{d-1}\\
 & \le & \frac{\pi^{\frac{d}{2}}}{\Gamma\left(\frac{d}{2}+1\right)}
(r+\sqrt{d})^d\ .
\end{eqnarray*}
Now if $d(\ux)\ge 2 B_r$, at least half of the $d(\ux)$
neighbors of $\ux$ are at a distance greater than $r$ from $\ux$.
This would imply
\[
e^{-\frac{\alpha}{2}\sum_{\uy|\{\ux,\uy\}\in\T}|\ux-\uy|}
\le e^{-\frac{\alpha r}{4}d(\ux)}\ .
\]
Let us first suppose that
\[
d(\ux)\ge
\frac{2\pi^{\frac{d}{2}}d^{\frac{d}{2}}}{\Gamma\left(\frac{d}{2}+1\right)}\ .
\]
Letting
\[
r_{\rm max}=\left\{
\frac{\Gamma\left(\frac{d}{2}+1\right)}{2\pi^{\frac{d}{2}}}
d(\ux)\right\}^{\frac{1}{d}}-\sqrt{d}\ ,
\]
we see that $r_{\rm max}\ge 0$
and $r=r_{\rm max}$ satisfies the hypothesis $d(\ux)\ge 2 B_r$.
Consequently, it follows that
\[
e^{-\frac{\alpha}{2}\sum_{\uy|\{\ux,\uy\}\in\T}|\ux-\uy|}
\le
\exp\left[
-\frac{\alpha d(\ux)}{4}
\left(\left(
\frac{\Gamma\left(\frac{d}{2}+1\right)}{2\pi^{\frac{d}{2}}}
d(\ux)
\right)^{\frac{1}{d}}-\sqrt{d}\right)
\right]\ .
\]
Using the trivial inequality $d(\ux)!\le d(\ux)^{d(\ux)}$ and the
definition
of $K_{2,2}$ which clearly is finite when $\alpha>0$, we have
\[
d(\ux)!\ e^{-\frac{\alpha}{2}\sum_{\uy|\{\ux,\uy\}\in\T}|\ux-\uy|}
\le K_{2,2}\ .
\]
In the second case where
\[
d(\ux)<
\frac{2\pi^{\frac{d}{2}}d^{\frac{d}{2}}}{\Gamma\left(\frac{d}{2}+1\right)}
\]
then (\ref{xandfriends})
is trivially bounded by $K_{2,1}$.
\qed

Since the $d(\ux)!$ to be bounded in (\ref{rhobound})
appear at the power $\frac{5}{2}$
and
we want to use only a fraction, say half, of the tree decay;
we will use the previous
lemma
with $\alpha=\frac{\mu_0}{5}$ so that $\frac{5}{2}\alpha=\frac{\mu_0}{2}$.

We now turn our attention to the $(m_0(\ux)+m_0^*(\ux))!^{\frac{1}{2}}$
in (\ref{rhobound}).
Let
\begin{eqnarray*}
c_I(\ux) & = & \left|\{i\in I| \ux_i=\ux\}\right|\ ,\\
b(\ux) & = & 4\bbone_{\{\ux\in\U\}}\ ,\\
v_\uy(\ux) & = & \left|\{j,1\le j\le k| \uy_j=\ux\}\right|\ .
\end{eqnarray*}
Then
\begin{eqnarray*}
\prod_{\ux\in R}
\left(m_0(\ux)+m_0^*(\ux)
\right)!^{\frac{1}{2}} & = &
\prod_{\ux\in R}
\left(c_I(\ux)+b(\ux)+4v_\uy(\ux)
\right)!^{\frac{1}{2}}\\
 & \le & \prod_{\ux\in R}
\left\{
c_I(\ux)!b(\ux)! v_\uy(\ux)!^4\times
6^{c_I(\ux)+b(\ux)+4v_\uy(\ux)}
\right\}^{\frac{1}{2}}\ .
\end{eqnarray*}
That is
\begin{equation}
\prod_{\ux\in R}
\left(m_0(\ux)+m_0^*(\ux)\right)!^{\frac{1}{2}}
\le
6^{\frac{1}{2}|I|+2|R|+2k}\times 24^{\frac{1}{2}|R|}\times
\prod_{\ux\in R} c_I(\ux)!^{\frac{1}{2}}\times
\prod_{\ux\in R} v_\uy(\ux)!^2
\label{mfactclean}
\end{equation}
where we used $|\U|\le |R|$.

We now simplify the bound (\ref{rhobound}) by using Lemma \ref{volumeeffect}
with $\alpha=\frac{\mu_0}{5}$, as well as
(\ref{mfactclean}).
We also bound the sum over the no longer needed $\U$
by $2^{|R|}$, and we bound $k$ by $M$ in the various exponents where
it appears.
After some cleaning up, we therefore get
\[
\left|
\left(\frac{d}{du}\right)^{M}
\tilde{\zeta}(R,I,u\l)
\right|\le
\left(\frac{\l}{4}\right)^{\max\{|R|-|I|,M\}}\
K_3^{|R|}K_4^{|I|} K_5^{M}
\ \bbone\left\{
\begin{array}{c}
\forall i\in I\\
\ux_i\in R
\end{array}
\right\}
\]
\[
\sum\limits_{k=0}^M
\frac{M!}{k!}
\times
\prod_{\ux\in R} c_I(\ux)!^{\frac{1}{2}}
\times
\left[
\sum_{\uy_1,\ldots,\uy_k\in R}
\ \prod_{\ux\in R} v_\uy(\ux)!^2
\right]
\times
\left[
\sum_{{\T\leadsto R}\atop{\T\ {\rm tree}}}
e^{-\frac{\mu_0}{2}\sum_{\{\ux,\uy\}\in\T} |\ux-\uy|}
\right]
\]
where
\begin{eqnarray*}
K_3 & = & 2^{16}3^{2}e^{10}\sqrt{6}K_1^5 K_2^{\frac{5}{2}}\ , \\
K_4 & = & 2e\sqrt{6K_1}\ ,\\
K_5 & = & 2^6 3^2 e^4K_1^2\ .
\end{eqnarray*}
Note that $K_2$ in Lemma \ref{volumeeffect}
is such that $K_2\ge K_{2,1}\ge 1$
and therefore $K_3,K_4,K_5$ all are $\ge 1$.

The sums over the positions $\uy_1,\ldots,\uy_k$ of the vertices
created
by the additional Taylor expansion, as well as over the positions
$\ux_2,\ldots,\ux_n$ in the $l^1$-clustering bound will be done thanks
to Lemma \ref{suminpoly}.

\begin{Remark}
We will use this lemma with both $\beta=2$ and $\beta=\frac{1}{2}$
in order to sum over the $\uy_j$ and $\ux_i$ respectively.
However the $\beta$ versus $1$ dichotomy prevents us
from compensating the factorials at the power $2$ by the ones at the
power $\frac{1}{2}$. This is the main bottleneck we found on the way to
Conjecture \ref{conj}.
\end{Remark}

Thanks to Lemma \ref{suminpoly}, the sum over the $\uy$'s is bounded
by $2^{|R|+k-1} k!^2$.
We then use the coarse bounds
\[
\sum\limits_{k=0}^{M} k!\  M!\ 2^{|R|+k-1} \le M!^2\ 2^{|R|+M}\ (M+1)
\le  M!^2\ 2^{|R|+2M}\ .
\]

As a result we have the desired bound which is summarized
in the following proposition.

\begin{Proposition}
Suppose the function $J:\Z^d\rightarrow \mathbb{C}$
satisfies $K_0=1$, where $K_0$ is the quantity defined in Lemma
\ref{decaylemma}.
Suppose $M$ is a nonnegative integer, $R$ is a polymer in $\L$, $u$
belongs to $[0,1]$ and $0<\l\le 4$. Then, we have the estimate
\[
\left|
\left(\frac{d}{du}\right)^{M}
\tilde{\zeta}(R,I,u\l)
\right|\le
\left(\frac{\l}{4}\right)^{\max\{|R|-|I|,M\}}\
(2K_3)^{|R|}K_4^{|I|} (4K_5)^{M} M!^2
\]
\begin{equation}
\times
\bbone\left\{
\begin{array}{c}
\forall i\in I\\
\ux_i\in R
\end{array}
\right\}
\times
\prod_{\ux\in R} c_I(\ux)!^{\frac{1}{2}}
\times
\left[
\sum_{{\T\leadsto R}\atop{\T\ {\rm tree}}}
e^{-\frac{\mu_0}{2}\sum_{\{\ux,\uy\}\in\T} |\ux-\uy|}
\right]
\label{singlepolybd}
\end{equation}
where right derivatives are meant when $u=0$, and left derivatives
when $u=1$.
\end{Proposition}

\subsubsection{The convergence criterion}\label{cvcritsection}
We now need to address the convergence criterion (\ref{cvcriterionzeta})
of Proposition \ref{cqftcorrseries}. Take any fixed site $\uz\in\L$,
we then need to bound the quantity
\[
Q=\sum_{R\in\mathbf{P}(\L)}
\bbone\{\uz\in R\} |\tilde{\zeta}(R,\emptyset,\l)| 2^{|R|}\ .
\]
The simple case $I=\emptyset$, $u=1$, $M=0$, of the basic raw estimate
(\ref{singlepolybd}) gives
\[
Q\le \sum_{R\in\mathbf{P}(\L)}
\bbone\{\uz\in R\}
\left(
\l  K_3
\right)^{|R|}
\sum_{{\T\leadsto R}\atop{\T\ {\rm tree}}}
e^{-\frac{\mu_0}{2}\sum_{\{\ux,\uy\}\in\T} |\ux-\uy|}\ .
\]
We now proceed as in \S\ref{largemasssection} and condition the sum on $m=|R|$
and also introduce
the
identity (\ref{relabeling}), in order to eliminate $R$.
The sum over the locations of the labeled sites gives
a factor $K_6^{m-1}$ using (\ref{expdecayeq})
and letting $K_6=K_1\left(d,\frac{\mu_0}{2}\right)$.
The sum over trees $\mathfrak{t}$ on the set $[m]$, this time
without the $d_{\mathfrak{t}}(i)!$'s is simply bounded by $m^{m-2}\le (m-2)! e^{m}$
by the coarser version of Cayley's Theorem~\cite[Prop. 5.3.2]{Stanley2}.
Note that one has to treat separately $m\ge 2$ and $m=1$.
The result is easily seen to again be a geometric series bound, namely
\[
Q\le \sum_{m\ge 1}
\left(
\l e K_3 K_6
\right)^m\ .
\]
Therefore if $\l\le \frac{1}{3e K_3K_6}<4$,
which we assume from now on,
then we have
\[
||\tilde{\zeta}(\cdot,\emptyset,\l)||\le \frac{1}{2}
\]
and the criterion is satisfied.

\subsubsection{Justification of the term by term differentiation}
We now address the issue of term by term differentiation leading
to the series expression (\ref{termbyterm}).
Recall that as a corrolary of the mean value theorem and Lebesgue
dominated convergence, the equation
\[
\left(\frac{d}{du}
\right)^N
\<\psi^{\sharp_1}(\ux_1),\ldots,\psi^{\sharp_n}(\ux_n)\>_{\L,u\l}^{\rm T}
=
\sum_{p\ge 1}\frac{1}{p!}
\sum_{{R_1,\ldots,R_p\in\mathbf{P}(\L)}\atop{I_1,\ldots,I_p\subset[n]}}
\phi^{\rm T}(R_1,\ldots,R_p)
\]
\begin{equation}
\bbone\left\{
\begin{array}{c}
I_q\ {\rm disjoint}\\
\cup I_q=[n]
\end{array}
\right\}
\sum_{N_1+\cdots+N_p=N}\frac{N!}{N_1!\ldots N_p!}
\prod_{q=1}^p
\left(\frac{d}{du}\right)^{N_q}
\tilde{\zeta}(R_q,I_q,u\l)
\label{tbyteq}
\end{equation}
will be established for $u\in [0,1]$
provided we can find majorants $G(R,I,\l,k)$ which are uniform in $u$
such that
\[
\left|
\left(\frac{d}{du}\right)^{k} \tilde{\zeta}(R,I,u\l)
\right|\le G(R,I,\l,k)
\]
for any $u\in [0,1]$, and such that for any integer $M$, $1\le M\le N$,
one has
\[
\sum_{p\ge 1}\frac{1}{p!}
\sum_{{R_1,\ldots,R_p\in\mathbf{P}(\L)}\atop{I_1,\ldots,I_p\subset[n]}}
|\phi^{\rm T}(R_1,\ldots,R_p)|
\bbone\left\{
\begin{array}{c}
I_q\ {\rm disjoint}\\
\cup I_q=[n]
\end{array}
\right\}
\]
\[
\sum_{M_1+\cdots+M_p=M}\frac{M!}{M_1!\ldots M_p!}
\prod_{q=1}^p G(R_q, I_q,\l, M_q)
< +\infty\ .
\]
We will take the majorants $G(R,I,\l,k)$ provided by the right-hand
side of (\ref{singlepolybd}). We therefore have to show the finiteness
of
\[
\mathcal{Q}=
\sum_{p\ge 1}\frac{1}{p!}
\sum_{{R_1,\ldots,R_p\in\mathbf{P}(\L)}\atop{I_1,\ldots,I_p\subset[n]}}
|\phi^{\rm T}(R_1,\ldots,R_p)|
\bbone\left\{
\begin{array}{c}
I_q\ {\rm disjoint}\\
\cup I_q=[n]
\end{array}
\right\}
\]
\[
\sum_{M_1+\cdots+M_p=M}\frac{M!}{M_1!\ldots M_p!}
\prod_{q=1}^p
\left[
\left(\frac{\l}{4}\right)^{\max\{|R_q|-|I_q|,M_q\}}\
(2K_3)^{|R_q|}K_4^{|I_q|} (4K_5)^{M_q} M_q!^2
\right.
\]
\[
\left.
\times
\bbone\left\{
\begin{array}{c}
\forall i\in I_q\\
\ux_i\in R_q
\end{array}
\right\}
\times
\prod_{\ux\in R_q} c_{I_q}(\ux)!^{\frac{1}{2}}
\times
\left[
\sum_{{\T_q\leadsto R_q}\atop{\T_q\ {\rm tree}}}
e^{-\frac{\mu_0}{2}\sum_{\{\ux,\uy\}\in\T_q} |\ux-\uy|}
\right]
\right]\ .
\]
Since the only issue is finiteness, we will use very coarse bounds
for $\mathcal{Q}$.
We write
\[
\left(\frac{\l}{4}\right)^{\max\{|R_q|-|I_q|,M_q\}}\le
\left(\frac{\l}{4}\right)^{|R_q|-|I_q|}
\]
as well as
\[
\prod_{q=1}^{p}
\prod_{\ux\in R_q} c_{I_q}(\ux)!^{\frac{1}{2}}\le n!^{\frac{1}{2}}\ .
\]
We also bound the characteristic functions by
\[
\bbone\left\{
\begin{array}{c}
I_q\ {\rm disjoint}\\
\cup I_q=[n]
\end{array}
\right\}
\prod_{q=1}^{p}
\bbone\left\{
\begin{array}{c}
\forall i\in I_q\\
\ux_i\in R_q
\end{array}
\right\}
\le
\bbone\left\{
\begin{array}{c}
I_q\ {\rm disjoint}\\
\cup I_q=[n]
\end{array}
\right\}
\bbone\{
\ux_1\in\cup_{q=1}^{p} R_q
\}\ .
\]
We finally use
\begin{equation}
\sum_{M_1+\cdots+M_p=M} M_1!\ldots M_p! \le M!\ 2^{M+p}
\label{summs}
\end{equation}
and
\[
\sum_{I_1,\ldots,I_p\subset [n]}
\bbone\left\{
\begin{array}{c}
I_q\ {\rm disjoint}\\
\cup I_q=[n]
\end{array}
\right\}
=p^n\le n! e^p
\]
and therefore get
\[
\mathcal{Q}\le
n!^{\frac{3}{2}}
\left(\frac{4 K_4}{\l}\right)^{n}
M!^2 (8K_5)^M
\]
\[
\sum_{p\ge 1}\frac{1}{p!}
\sum_{R_1,\ldots,R_p\in \mathbf{P}(\L)}
\bbone\{\ux_1\in\cup_{q=1}^p R_q\}
|\phi^{\rm T}(R_1,\ldots,R_p)|.
\rho(R_1)\ldots\rho(R_p)
\]
with
\[
\rho(R)= 2e
\left(\frac{\l K_3}{2}\right)^{|R|}
\sum_{{\T\leadsto R}\atop{\T\ {\rm tree}}}
e^{-\frac{\mu_0}{2}\sum_{\{\ux,\uy\}\in\T} |\ux-\uy|}\ .
\]
Now using the same argument as in \S\ref{cvcritsection}
we see that
\[
||\rho||\le 2e \sum_{m\ge 1}
\left(\l e K_3 K_6\right)^{m}\ .
\]
As a result, we will have
$||\rho||\le 1$
provided
\[
\l\le \frac{1}{(1+2e) e K_3 K_6}<\frac{1}{3 e K_3K_6}
\]
which we now assume.
Finally, Lemma \ref{pinsumlemma}
shows that $\mathcal{Q}$ is finite, and therefore
(\ref{tbyteq}) as well as (\ref{termbyterm}) are justified.

\subsubsection{The clustering estimate}\label{finalclusterbound}
We now, under the hypotheses $n\ge 2(N+1)$,
$K_0=1$ and $0<\l\le \frac{1}{(1+2e) e K_3 K_6}$,
come to the clustering estimate proper, i.e., the bound on
\[
\mathfrak{C}=
\sum_{\ux_1,\ldots,\ux_n\in\L}
\bbone\{\ux_1=\zero\}\
|\<\psi^{\sharp_1}(\ux_1),\ldots,\psi^{\sharp_n}(\ux_n)\>_{\L,\l}^{\rm T}|\ .
\]
Using (\ref{termbyterm}),
we can write
\[
\mathfrak{C}\le
\sum_{\ux_1,\ldots,\ux_n\in\L}
\bbone\{\ux_1=\zero\}
\int\limits_0^1 \ du\ \frac{(1-u)^{N-1}}{(N-1)!}
\sum_{p\ge 1}\frac{1}{p!}
\sum_{{R_1,\ldots,R_p\in\mathbf{P}(\L)}\atop{I_1,\ldots,I_p\subset[n]}}
|\phi^{\rm T}(R_1,\ldots,R_p)|
\]
\[
\bbone\left\{
\begin{array}{c}
I_q\ {\rm disjoint}\\
\cup I_q=[n]
\end{array}
\right\}
\sum_{N_1+\cdots+N_p=N}\frac{N!}{N_1!\ldots N_p!}
\prod_{q=1}^p
\left|\left(\frac{d}{du}\right)^{N_q}
\tilde{\zeta}(R_q,I_q,u\l)\right|\ .
\]
Inserting the bound (\ref{singlepolybd}), performing the $u$ integral,
and tidying the resulting inequality, one obtains
\begin{eqnarray*}
\mathfrak{C} & \le & \left(\l K_5\right)^N K_4^n
\sum_{\ux_1,\ldots,\ux_n\in\L}
\bbone\{\ux_1=\zero\}
\sum_{p\ge 1}\frac{1}{p!}
\sum_{{R_1,\ldots,R_p\in\mathbf{P}(\L)}\atop{I_1,\ldots,I_p\subset[n]}}\\
 & & |\phi^{\rm T}(R_1,\ldots,R_p)|
\bbone\left\{
\begin{array}{c}
I_q\ {\rm disjoint}\\
\cup I_q=[n]
\end{array}
\right\}
\sum_{N_1+\cdots+N_p=N} N_1!\ldots N_p!\\
 & & \prod_{q=1}^{p}
\left[
\left(\frac{\l}{4}\right)^{\max\{|R_q|-|I_q|-N_q,0\}}
(2K_3)^{|R_q|}
\times
\bbone\left\{
\begin{array}{c}
\forall i\in I_q\\
\ux_i\in R_q
\end{array}
\right\}
\right.\\
 & & \left.
\times\prod_{\ux\in R_q}c_{I_q}(\ux)!^{\frac{1}{2}}
\times\left\{
\sum_{{\T_q\leadsto R_q}\atop{\T_q\ {\rm tree}}}
e^{-\frac{\mu_0}{2}\sum_{\{\ux,\uy\}\in\T_q} |\ux-\uy|}
\right\}
\right]\ .
\end{eqnarray*}
We now proceed as in \S\ref{largemasssection}
and push $\bbone\{\ux_1=\zero\}$ through the sums over
$p$, the $R_q$'s and the $I_q$'s, before bounding it by the coarser condition
$\bbone\{\zero\in\cup_{q=1}^p R_q\}$.
We likewise push the sums over the $\ux_i$'s
inside the appropriate
bracket factor.
The sums over the source localizations $\ux_i$ are then performed
using Lemma \ref{suminpoly} and yield the same bound as
(\ref{localsumhalf}).
Hence,
\[
\mathfrak{C}\le
\left(\l K_5\right)^N (2K_4)^n
\sum_{p\ge 1}\frac{1}{p!}
\sum_{{R_1,\ldots,R_p\in\mathbf{P}(\L)}\atop{I_1,\ldots,I_p\subset[n]}}
\bbone\{\zero\in\cup_{q=1}^p R_q\}
\]
\[
|\phi^{\rm T}(R_1,\ldots,R_p)|
\bbone\left\{
\begin{array}{c}
I_q\ {\rm disjoint}\\
\cup I_q=[n]
\end{array}
\right\}
|I_1|!\ldots|I_p|!\times
\sum_{N_1+\cdots+N_p=N} N_1!\ldots N_p!
\]
\[
\prod_{q=1}^{p}
\left[
\left(\frac{\l}{4}\right)^{\max\{|R_q|-|I_q|-N_q,0\}}
(4K_3)^{|R_q|}
\times\left\{
\sum_{{\T_q\leadsto R_q}\atop{\T_q\ {\rm tree}}}
e^{-\frac{\mu_0}{2}\sum_{\{\ux,\uy\}\in\T_q} |\ux-\uy|}
\right\}
\right]\ .
\]
We now introduce a constant $\gamma$, $0<\gamma\le 1$, to be fine-tuned
shortly; and we suppose that $\l$ satisfies the extra
hypothesis $0<\frac{\l}{4}\le\gamma$.
We can now use the estimate
\[
\left(\frac{\l}{4}\right)^{\max\{|R_q|-|I_q|-N_q,0\}}
\le \gamma^{\max\{|R_q|-|I_q|-N_q,0\}}\le \gamma^{|R_q|-|I_q|-N_q}\ .
\]
We also use the previously derived inequalities (\ref{sumis})
and (\ref{summs}) which yield
\begin{eqnarray*}
\lefteqn{
\mathfrak{C}\le
\left(\frac{2\l K_5}{\gamma}\right)^N
(\frac{4K_4}{\gamma})^n N! n!} & & \\
 & & \times \sum_{p\ge 1}\frac{1}{p!}
\sum_{{R_1,\ldots,R_p\in\mathbf{P}(\L)}\atop{I_1,\ldots,I_p\subset[n]}}
\bbone\{\zero\in\cup_{q=1}^p R_q\}
|\phi^{\rm T}(R_1,\ldots,R_p)|
\prod_{q=1}^{p}\varpi(R_q)
\end{eqnarray*}
with
\[
\varpi(R)=4.(4\gamma K_3)^{|R|}
\sum_{{\T\leadsto R}\atop{\T\ {\rm tree}}}
e^{-\frac{\mu_0}{2}\sum_{\{\ux,\uy\}\in\T} |\ux-\uy|}\ .
\]
Now, again as in \S\ref{cvcritsection}, we
have
\[
||\varpi||\le 4 \sum_{m\ge 1}\left(8e\gamma  K_3 K_6\right)^{m}\ .
\]
Thus we will have
$||\varpi||\le 1$ as soon as
\[
\gamma\le\frac{1}{40 e K_3 K_6}\ .
\]
We therefore choose $\gamma=\frac{1}{40 e K_3 K_6}$.
Since we have by hypothesis that
\[
0<\l\le 4\gamma=
\frac{1}{10 e K_3 K_6}< \frac{1}{(1+2e)e K_3 K_6}
< \frac{1}{3 e K_3 K_6}<4
\]
we checked the validity of every condition we needed to check and
we can use
Lemma \ref{pinsumlemma}
and conclude
\[
\mathfrak{C}\le
\left(\frac{2\l K_5}{\gamma}\right)^N (\frac{4K_4}{\gamma})^n N! n!\ .
\]
In other words we have the desired clustering bound
\begin{eqnarray*}
\lefteqn{
\sum_{\ux_2,\ldots,\ux_n\in\L}
|\<\psi^{\sharp_1}(\zero),\psi^{\sharp_2}(\ux_2),\ldots,\psi^{\sharp_n}(\ux_n)
\>_\L^{\rm T}|\le} & & \\
 & & \l^N\ n!\ N!\ (80 e K_3 K_5 K_6)^N\ (160 e K_3 K_4 K_6)^n\ .
\end{eqnarray*}

We will now get rid of the restriction to $K_0=1$
by a simple scaling transformation on the field variables $\psi$.
Indeed if one does the change of variable $\psi=\eta \psi'$, for some
$\eta>0$,
in the original model, one easily sees that the $n$-point truncated
correlations of the $\psi$ fields becomes $\eta^n$
times the analogous correlation for the $\psi'$ fields.
The latter are sampled according to the measure
corresponding to the input parameters
$\eta^2 J$, $\eta ^4\l$ instead of the original function $J$ and coupling $\l$
respectively.
Now one can go back through the definitions of the various constants
$\mu_0$, $\alpha$, $K_1$, $K_2$, $K_3$, $K_4$, $K_5$ and $K_6$,
and easily check that they are invariant if one multiplies
the $J$ function by a nonzero scalar $\eta^2$. On the other hand
$K_0=\frac{J(\zero)}{J_{\neq}(J(\zero)-J_{\neq})}$
gets multiplied by $\eta^{-2}$.
Therefore one can make the new $K_0=1$
by choosing $\eta=\sqrt{\frac{J(\zero)}{J_{\neq}(J(\zero)-J_{\neq})}}$.
Thus we have proved that if
\[
0<\l\le\frac{J_{\neq}^2(J(\zero)-J_{\neq})^2}{10 e K_3 K_6 J(\zero)^2}\ ;
\]
then, uniformly in $\L\subset\Z^d$, and $n$ even satisfying $n\ge 2(N+1)$,
we have
\[
\sum_{\ux_2,\ldots,\ux_n\in\L}
|\<\psi^{\sharp_1}(\zero),\psi^{\sharp_2}(\ux_2),\ldots,\psi^{\sharp_n}(\ux_n)
\>_\L^{\rm T}|\le
c_1(N,J)\times c_2(J)^n\times \l^N\times n!
\]
where
\[
c_1(N,J)=N!\left(
\frac{80 e K_3 K_5 K_6 J(\zero)^2}{J_{\neq}^2(J(\zero)-J_{\neq})^2}
\right)^N
\]
and
\[
c_2(J)=160 e K_3 K_4 K_6
\sqrt{\frac{J(\zero)}{J_{\neq}(J(\zero)-J_{\neq})}}\ .
\]
This completes the proof of Theorem \ref{smalllambdathm}, i.e.,
the clustering estimate for
$n$-point functions with $n\ge 4$.
\qed

The proof of Theorem \ref{twoptfunction} is exactly the same
as the one given above for Theorem \ref{smalllambdathm} with the choices
$n=2$ and $N=1$.
The only difference is that the right-hand side of the starting point equation
(\ref{termbyterm}) now expresses the Taylor remainder
\[
\<\psi^{\sharp_1}(\ux_1),\psi^{\sharp_2}(\ux_2)\>_{\L,\l}^{\rm T}
-\<\psi^{\sharp_1}(\ux_1),\psi^{\sharp_2}(\ux_2)\>_{\L,0}^{\rm T}
\]
instead of the full 2-point function.

\bigskip
\noindent{\bf Acknowledgements:}
{\small
We thank D.~Brydges, J.~Imbrie, J.~Magnen, A.~Sokal,
H.~Spohn and D.~Wagner for useful discussions or correspondence.
We wish to thank the Isaac Newton Institute for Mathematical Sciences,
University of Cambridge, for generous support during the programme on
Combinatorics and Statistical Mechanics (January--June 2008),
where this work was initiated. We also thank the organizers of this program
P.~Cameron, B.~Jackson, A.~Scott, A.~Sokal and D.~Wagner
for their invitation.
The first author also thanks the Dean of Arts and Sciences
of the University of Virginia for permission to attend this program
for several extended periods.
The second author aknowledges the support of the Conselho Nacional
de Desenvolvimento Cient\'{\i}fico
e Tecnol\'ogico (CNPq), and the Funda\c{c}\~{a}o de Amparo \`a Pesquisa do
Estado de Minas Gerais (FAPEMIG).}


\begin{thebibliography}{99}

\bibitem{AbdesselamSLC}
A.~Abdesselam. Feynman diagrams in algebraic combinatorics.
S\'em. Lothar. Combin. 49 (2002/04), Art. B49c, 45 pp. (electronic).

\bibitem{AR1}
A.~Abdesselam and V.~Rivasseau.
Trees, forests and jungles: a botanical garden for cluster expansions.
Constructive physics (Palaiseau, 1994), 7--36,
Lecture Notes in Phys., 446. Berlin: Springer, 1995.

\bibitem{AR2}
A.~Abdesselam and V.~Rivasseau.
An explicit large versus small field multiscale cluster expansion.
Rev. Math. Phys. 9 (1997), no. 2, 123--199.

\bibitem{AdamsF}
R.~A.~Adams and J.~J.~F.~Fournier.
Sobolev Spaces. Second edition. New York: Academic Press, 2003.

\bibitem{AndrewsAR}
G.~E.~Andrews, R.~Askey and R.~Roy.
Special functions.
Encyclopedia of Mathematics and its Applications, 71.
Cambridge University Press, Cambridge, 1999.

\bibitem{Aneetal}
C.~An\'e, S.~Blach\`ere, D.~Chafa\"{\i}, P.~Foug\`eres, I.~Gentil,
F.~Malrieu, C.~Roberto, and G.~Scheffer.
Sur les in\'egalit\'es de Sobolev logarithmiques.
With a preface by Dominique Bakry and Michel Ledoux.
Panoramas et Synth\`eses, 10. Soc. Math. France, Paris, 2000.

\bibitem{BachJS}
V.~Bach, T.~Jecko and J.~Sj\"{o}strand.
Correlation asymptotics of classical lattice spin systems with
nonconvex Hamilton function at low temperature.
Ann. Henri Poincar\'e 1 (2000), no. 1, 59--100.

\bibitem{BachM1}
V.~Bach and J.~S.~M{\o}ller.
Correlation at low temperature. I. Exponential decay.
J. Funct. Anal. 203 (2003), no. 1, 93--148.

\bibitem{BIJ}
T.~Balaban, J.~Z.~Imbrie and A.~Jaffe.
Effective action and cluster properties of the Abelian Higgs model.
Comm. Math. Phys. 114 (1988), 257--315.

\bibitem{BattleF}
G.~A.~Battle and P.~Federbush.
A note on cluster expansions, tree graph identities, extra $1/N!$ factors!
Lett. Math. Phys. 8 (1984), no. 1, 55--57.

\bibitem{Birnbaum}
Z.~W.~Birnbaum.
An inequality for Mill's ratio.
Annals of Math. Stat. 13 (1942), no. 2, 245--246.

\bibitem{Brydges}
D.~C.~Brydges.
A short course on cluster expansions.
Ph\'enom\`enes critiques, syst\`emes al\'eatoires, th\'eories de jauge,
Part I, II (Les Houches, 1984),  129--183, North-Holland, Amsterdam, 1986.

\bibitem{BDH}
D.~Brydges, J.~Dimock and T.~R.~Hurd.
The short distance behavior of $(\phi\sp 4)\sb 3$.
Comm. Math. Phys. 172 (1995), 143--186.

\bibitem{BF}
D.~C.~Brydges and P.~Federbush.
A new form of the Mayer expansion in classical statistical mechanics.
J. Math. Phys. 19 (1978), no. 10, 2064--2067.

\bibitem{BK}
D.~Brydges and T.~Kennedy.
Mayer expansions and the Hamilton-Jacobi equation.
J. Stat. Phys. 48 (1987), 19

\bibitem{BM}
D.~Brydges and P.~Martin.
Coulomb systems at low density: a review.
J. Stat. Phys. 96 (1999), 1163--1330.

\bibitem{Cammarota}
C.~Cammarota.
Decay of correlations for infinite range interactions in unbounded
spin systems.
Comm. Math. Phys. 85 (1982), no. 4, 517--528.

\bibitem{Constantinescu}
F.~Constantinescu.
Analyticity in the coupling constant of the $\lambda P(\varphi )$
lattice theory.
J. Math. Phys. 21 (1980), no. 8, 2278--2281.

\bibitem{DIS}
M.~Duneau, D.~Iagolnitzer and B.~Souillard.
Decrease properties of truncated correlation
functions and analyticity properties for
classical lattices and continuous systems.
Comm. Math. Phys. 31 (1973), 191--208.

\bibitem{EMS}
J.-P.~Eckmann, J.~Magnen and R.~S\'en\'eor.
Decay properties and Borel summability for the Schwinger functions
in $P(\phi)\sb{2}$ theories.
Comm. Math. Phys. 39 (1974/75), 251--271.

\bibitem{FernandezP}
R.~Fern\'andez and A.~Procacci.
Cluster expansion for abstract polymer models. New bounds from an
old approach.
Comm. Math. Phys. 274 (2007), no. 1, 123--140.

\bibitem{GJ}
J.~Glimm and A.~Jaffe.
Quantum physics. A functional integral point of view. Second edition.
New York: Springer, 1987.

\bibitem{GJS1}
J.~Glimm, A.~Jaffe and T.~Spencer.
The Wightman axioms and particle structure in the $P(\phi)_2$ quantum
field model.
Annals of Math. 100 (1974), 585--632.

\bibitem{GJS2}
J.~Glimm, A.~Jaffe and T.~Spencer.
The particle structure of the weakly coupled $P(\phi)_2$ model and
other applications of high temperature expansions, Part II: The
cluster expansion. Constuctive quantum field theory, Erice 1973,
eds. G.~Velo and A.~Wightman, Lecture notes in Phys. 25.
New York: Springer 1973.

\bibitem{Knuth}
R.~L.~Graham, D.~E.~Knuth and O.~Patashnik.
Concrete mathematics.
A foundation for computer science.
Addison-Wesley Publishing Company, Advanced Book Program, Reading, MA, 1989.

\bibitem{Gross}
L.~Gross.
Decay of correlations in classical lattice models at high temperature.
Comm. Math. Phys. 68 (1979), no. 1, 9--27.

\bibitem{GruberK}
C.~Gruber and H.~Kunz.
General properties of polymer systems.
Comm. Math. Phys. 22 (1971), 133--161.

\bibitem{GuionnetZ}
A.~Guionnet and B.~Zegarlinski.
Lectures on logarithmic Sobolev inequalities.
S\'eminaire de Probabilit\'es, XXXVI,  1--134,
Lecture Notes in Math., 1801, Springer, Berlin, 2003.

\bibitem{Helffer}
B.~Helffer.
Semiclassical analysis, Witten Laplacians, and statistical mechanics.
Series in Partial Differential Equations and Applications, 1.
World Scientific Publishing Co., Inc., River Edge, NJ, 2002.

\bibitem{HelfferS}
B.~Helffer and J.~Sj\"{o}strand.
On the correlation for Kac-like models in the convex case.
J. Statist. Phys. 74 (1994), no. 1-2, 349--409.

\bibitem{IsraelN}
R.~B.~Israel and C.~R.~Nappi.
Exponential clustering for long-range integer-spin systems.
Comm. Math. Phys. 68 (1979), no. 1, 29--37.

\bibitem{Isserlis}
L.~Isserlis.
On a Formula for the product-moment coefficient
of any order of a normal frequency distribution in any number of variables.
Biometrika 12 (1918), 134-139.

\bibitem{Kunz}
H.~Kunz.
Analyticity and clustering properties of unbounded spin systems.
Comm. Math. Phys. 59 (1978), no. 1, 53--69.

\bibitem{LeBellac}
M.~Le Bellac. Quantum and statistical field theory.
Translated from the 1988 French original by G. Barton and revised
by the author. The Clarendon Press, Oxford University Press,
New York, 1991.

\bibitem{Lo}
A.~Lo.
On the exponential decay of the $n$-point correlation
functions and the analyticity of the pressure.
J. Math. Phys. 48 (2007), no. 12, 123506.

\bibitem{LukkarinenS}
J.~Lukkarinen and H.~Spohn.
Not to normal order -- Notes on the kinetic limit for
weakly interacting quantum fluids.
Preprint arXiv:0807.5072v1 [math-ph], 2008.

\bibitem{LukkarinenS2}
J.~Lukkarinen and H.~Spohn.
Weakly nonlinear Schr\"{o}dinger equation with random initial data.
Preprint arXiv:0901.3283v1 [math-ph], 2009.

\bibitem{MackP}
G.~Mack and A.~Pordt.
Convergent perturbation expansions for Euclidean quantum field theory.
Comm. Math. Phys. 97 (1985), no. 1-2, 267--298.

\bibitem{MalyshevM}
V.~A.~Malyshev and R.~A.~Minlos.
Gibbs random fields.
Cluster expansions. Translated from the Russian by
R. Koteck\'y and P. Holick\'y.
Mathematics and its Applications (Soviet Series), 44.
Kluwer Academic Publishers Group, Dordrecht, 1991.

\bibitem{Matte}
O.~Matte.
Supersymmetric Dirichlet operators, spectral gaps, and correlations.
Ann. Henri Poincar\'e 7 (2006), no. 4, 731--780.

\bibitem{Pordt}
A.~Pordt.
Mayer expansions for Euclidean lattice field theory:
convergence properties and relation with perturbation theory.
Desy preprint 85-103, unpublished, 1985. Available at
\texttt{http://www-lib.kek.jp/top-e.html}

\bibitem{PdLS}
A.~Procacci, B.~N.~B.~de Lima and B.~Scoppola.
A remark on high temperature polymer expansion for lattice systems
with infinite range pair interactions.
Lett. Math. Phys. 45 (1998), no. 4, 303--322.

\bibitem{ProcacciS}
A.~Procacci and B.~Scoppola.
On decay of correlations for unbounded spin systems
with arbitrary boundary conditions.
J. Statist. Phys. 105 (2001), no. 3-4, 453--482.

\bibitem{Rivasseaubook}
V.~Rivasseau.
From perturbative to constructive renormalization.
Princeton Series in Physics. Princeton University Press, Princeton, NJ, 1991.

\bibitem{RuelleRMP}
D.~Ruelle.
Cluster property of the correlation functions of classical gases.
Rev. Modern Phys. 36 (1964), 580--584.

\bibitem{Ruelle}
D.~Ruelle.
Statistical mechanics: Rigorous results. W. A. Benjamin, Inc.,
New York-Amsterdam, 1969.

\bibitem{Salm}
M.~Salmhofer. Renormalization. An introduction.
Texts and Monographs in Physics. Springer-Verlag, Berlin, 1999.

\bibitem{Salm2}
M.~Salmhofer.
Clustering of fermionic truncated expectation values via functional
integration.
Preprint arXiv:0809.3517v1 [math-ph], 2008.

\bibitem{Simon}
B.~Simon.
The statistical mechanics of lattice gases. Vol. I.
Princeton Series in Physics. Princeton University Press,
Princeton, NJ, 1993.

\bibitem{Sjostrand1}
J.~Sj\"{o}strand.
Correlation asymptotics and Witten Laplacians.
Algebra i Analiz  8  (1996),  no. 1, 160--191;  translation in
St. Petersburg Math. J. 8 (1997), no. 1, 123--147.

\bibitem{Sjostrand2}
J.~Sj\"{o}strand.
Complete asymptotics for correlations of
Laplace integrals in the semi-classical limit.
M\'em. Soc. Math. France (N.S.) No. 83 (2000).

\bibitem{Sokal}
A.~Sokal.
Mean-field bounds and correlation inequalities.
J. Statist. Phys. 28 (1982), no. 3, 431--439.

\bibitem{Sokal2}
A.~D.~Sokal.
Bounds on the complex zeros of (di)chromatic polynomials and
Potts-model partition functions.
Combin. Probab. Comput. 10 (2001), no. 1, 41--77.

\bibitem{Stanley2}
R.~P.~Stanley.
Enumerative combinatorics. Vol. 2.
With a foreword by Gian-Carlo Rota and appendix 1 by Sergey Fomin.
Cambridge Studies in Advanced Mathematics, 62.
Cambridge University Press, Cambridge, 1999.

\bibitem{Wagner}
W.~Wagner.
Analyticity and Borel-summability of the perturbation
expansion for correlation functions of continuous spin systems.
Helv. Phys. Acta 54 (1981/82), no. 3, 341--363.

\bibitem{Wick}
G.~C.~Wick.
The evaluation of the collision matrix.
Physical Rev. (2) 80 (1950), 268--272.

\bibitem{Yoshida}
N.~Yoshida.
The log-Sobolev inequality for weakly coupled lattice fields.
Probab. Theory Related Fields 115 (1999), no. 1, 1--40.

\bibitem{Yoshida2}
N.~Yoshida.
The equivalence of the log-Sobolev inequality and a mixing
condition for unbounded spin systems on the lattice.
Ann. Inst. H. Poincar\'e Probab. Statist. 37 (2001), no. 2, 223--243.

\bibitem{Zegarlinski}
B.~Zegarlinski.
The strong decay to equilibrium for the stochastic dynamics of
unbounded spin systems on a lattice.
Comm. Math. Phys. 175 (1996), no. 2, 401--432.

\end{thebibliography}
\end{document}